\documentclass[manuscript]{aastex}
\usepackage{longtable}
\usepackage{graphicx}

\shorttitle{Mildly H-deficeint stars of Omega Cen}
\shortauthors{Hema & Pandey}

\begin{document}

\title{High resolution spectroscopy of the relatively 
hydrogen-poor \\ metal rich giants in the globular cluster 
$\omega$ Centauri }

\author{B. P. Hema\altaffilmark{1},  Gajendra Pandey\altaffilmark{1} and R. Srianand\altaffilmark{2}}

\affil{$^{1}$ Indian Institute of Astrophysics, Koramangala II Block, Bengaluru, Karnataka, India-560034 }

\affil{$^{2}$ Inter-University Centre for Astronomy and Astrophysics, Post Bag 4, Ganeshkhind, Pune 411007, India}

\email{hema@iiap.res.in}

\begin{abstract}

High-resolution optical spectra are analyzed  
for two of the four metal rich mildly hydrogen-poor or
helium-enhanced giants discovered by \citet{hema14} 
along with their comparison normal (hydrogen-rich) 
giants. The strengths of the MgH bands in the spectra 
of the program stars are analyzed for their 
derived stellar parameters. The observed spectra of the
sample (hydrogen-poor) stars  (LEID 39048 and LEID 34225) 
show weaker MgH bands unlike in the spectra of the 
normal comparison giants (LEID 61067 and LEID 32169). 
The magnesium abundance derived 
from MgH bands is less by 0.3 dex or more
for LEID 39048 and LEID 34225, than that derived from 
Mg\,{\sc i} lines. This difference, cannot be reconciled 
by making the changes to the stellar parameters within 
the uncertainties. This difference in the magnesium 
abundances derived from 
Mg\,{\sc i} lines and from the MgH band is unacceptable. 
This difference is attributed to the hydrogen-deficiency or
helium-enhancement in their atmospheres. These 
metal rich hydrogen-poor or helium-rich giants provide an 
important link to the evolution of the metal-rich sub 
population of $\omega$ Cen. These stars provide the
first direct spectroscopic evidence for the presence of
the He-enhancement in the metal rich giants of $\omega$ Cen.

\end{abstract}


\section{Introduction}

The brightest and the most massive Galactic globular 
cluster (GGC) $\omega$ Cen exhibits a large spread 
in metallicity ([Fe/H)] and other abundance 
anomalies among the cluster stars
\citep{johnson10, simpson13, sollima05} including 
the existence of multiple stellar population with 
the cluster dwarfs/giants having normal helium, 
and  enhanced helium or relatively hydrogen-poor 
(H-poor) atmospheres \citep{piotto05, villanova07, dupree13}.

The doubt is that these relatively H-poor or He-rich 
giants appear metal rich, than they really are, due 
to the lower continuum absorption. Note that the 
spectra of H-poor F- and G-type supergiants, R Coronae
Borealis (RCB) and the H-deficient carbon (HdC) stars,
appear metal rich when compared with the spectra of 
normal F- and G-type supergiants \citep{pandey04}.
This is ascribed to the lower continuum opacity in 
the atmosphere due to H-deficiency. Hence, these 
stars appear more metal rich than they actually are
\citep{sumangala11}. The origin and evolution of 
these enigmatic stars, RCB/HdC, are not yet clear 
due to their rarity and chemical peculiarity.

A low resolution optical spectroscopic survey 
was conducted among the giants of $\omega$ Cen 
for idetifying new R Coronae Borealis (RCB) stars 
\citep{hema14}. Identifying RCB stars in a globular 
cluster gives an idea of their position on the 
HR-diagram and is a potential clue to understand 
their origin and evolution.

\citet{hema14} survey resulted in the discovery 
of four mildly H-deficient giants in $\omega$ Cen. 
For two out of these four giants, high resolution 
spectra were obtained along with their 
comparison stars with similar stellar parameters 
as these two giants. 
In this paper we have conducted a detailed high resolution 
spectroscopic analyses for two mildly H-deficient giants
along with their comparison stars to confirm their 
H-deficiency or the He-enhancement.

\section{Observations and data reduction}

The high-resolution optical spectra  
were obtained using Southern African Large 
Telescope (SALT) $-$ high resolution spectrograph 
(HRS)\footnote{SALT HRS is a dual beam fibre-fed, 
white-pupil, echelle spectrohraph, employing
VPH gratings as cross dispersers.}. 
These spectra obtained with the SALT-HRS are having 
a resolving power, R ($\lambda$/$\Delta$$\lambda$) of 
40000. The spectra were obatined with both blue and 
red camera using 2K $\times$ 4K and 4K $\times$ 4K CCDs, 
respectively, spanning a spectral range of 
370$-$550 nm in the blue and 550$-$890 nm in the red.

The spectral reductions were carried out using the 
IRAF software. The traditional data reduction procedure 
including, bias subtraction, flat field correction, 
spectrum extraction, wavelength caliberation, etc. 
were followed.

The extracted and wavelength calibrated 1d spectra were 
continuum normalized. The region of the spectrum 
with maximum flux points and free of absorption lines 
were considered for continuum fitting with a smooth 
curve passing through these points. 
To improve the signal-to-noise, the observed  
spectra of the program stars were smoothed such that
the strength of the spectral lines are not altered.
The signal-to-noise ratio per pixel of the smoothed 
blue spectra of the program stars at about 5000\AA\ is 
$\sim$ 150 and for red spectra at about 
7000\AA\ is $\sim$ 200.
Since there is a overlap of wavelengths, 
the spectrum is continuous without gaps in the 
blue and red spectral range. The atlas of 
high-resolution spectrum of  Arcturus \citep{hinkle00}
was used as a reference for continuum fitting and 
also for identifying the spectral lines.

\section{Abundance Analysis}

In order to conduct a detailed abundance analysis, the 
equivalent widths for weak and strong lines of
several elements, that are clean and also 
not severely blended for both 
neutral and ionized states were meaured using 
different commands in IRAF.

The wavelength calibrated spectrum of a program star 
(LEID 39048) was used to measure the shift of the spectral lines from the 
rest wavelengths; \citet{hinkle00}'s atlas of Arcturus was used as reference. 
Adopting LEID 39048's spectrum as a template, the radial velocities 
and the uncertainties involved were determined for other program stars
using the task $fxcor$ in IRAF. Then the heliocentric corrections 
were applied to these radial velocities. The heliocentric velocities 
for the program stars are given in Table 1, and are in good agreement 
with the mean velocity of the cluster, $V_{r}$ = 233 km\,s$^{-1}$, 
with the dispersion ranging from 15 to 6 km\,s$^{-1}$ from 
center to outwards, respectively \citep{mayor97}.

For determination of stellar parameters and the 
elemental abundances, the LTE line analysis and spectrum 
synthesis code MOOG \citep{sneden73} and the ATLAS9 
\citep{kurucz98} plane parallel, line-blanketed
LTE model atmospheres were used. 

The microturbulence ($\xi_{t}$) is derived 
using Fe\,{\sc i} lines
having similar excitation potential and 
a range in equivalent width, weak to strong,
giving the same abundance. 
The effective temperature ($T_{\rm eff}$) is determined 
using the excitation balance of Fe\,{\sc i} lines having
a range in  lower excitation potential. 
The $T_{\rm eff}$ and $\xi_{t}$ were fixed iteratively.
The process was carried out untill both returned zero
slope. 

By adopting the determined $T_{\rm eff}$ 
and $\xi_{t}$, the surface gravity (log $g$) 
is derived. 
The surface gravity is fixed by demanding the same 
abundances from the lines of different ionization 
states of a species, known as ionization balance. 
The surface gravity is derived  
using the lines of Fe\,{\sc i}/Fe\,{\sc ii}, 
Ti\,{\sc i}/Ti\,{\sc ii}, and Sc\,{\sc i}/Sc\,{\sc ii}. 
Then the mean log $g$ was adopted.

Uncertainty on the $T_{\rm eff}$ and 
$\xi_{t}$  is estimated 
by changing the $T_{\rm eff}$ in steps of 25K and $\xi_{t}$
in steps of 0.05 km\,s$^{-1}$. The change in 
$T_{\rm eff}$ and $\xi_{t}$ and the corresponding change, in the  
mean abundances from the the mean abundances 
(of zero slope), of about 1$\sigma$ error 
on the the mean abundances (of zero slope) is obtained.
This change is adopted as the uncertainty on these parameters.  
The adopted $\Delta$$T_{\rm eff}$ = $\pm$50\,K and 
$\Delta$$\xi_{t}$ = $\pm$0.1\,km\,s$^{-1}$ (see Figure 2). 
The uncertainties on log $g$ is the standard deviation 
from the mean value of the log $g$ determined from 
different species, which is about $\pm$0.1 (cgs units). 
The adopted stellar parameters with the uncertainties 
are given in Table 1.

The log $g$ values for the program stars in \citet{hema14} were
derived using the standard relation: 

log(g$_{*}$)\,$=$\,0.40($M_{\rm bol.}$ $-$ $M_{\rm bol, \odot}$)
$+$ log($g_{\odot}$) $+$ 4(log($T$/$T_{\odot}$))
$+$ log($M$/$M_{\odot}$) --- Eq (1)

The bolometric correction to $M_{v}$ was applied 
using the relation given by \citet{alonso99}. The 
distance modulus for $\omega$ Cen, $(m-M)_{v}$ = 13.7,
and the mass of $\omega$ Cen 
red giants were assumed to be 0.8$M_{\odot}$ \citep{johnson10}.
Using the photometric temperatures derived from ${(J-H)_0}$, 
${(J-K)_0}$ and $(b-y)$ colours in Equation (1) the log $g$ values
 were derived.

The difference in the log $g$ values derived by us 
and those derived by \citet{johnson10} are within
$\pm$0.1 (cgs units). Hence, the uncertainty on 
the log $g$ values adopted by \citet{hema14} 
was about $\pm$0.1 (cgs units) (for details 
see \citet{hema14}).

Color-magnitude diagram for our program stars and 
for the sample red giants of \citet{johnson10} is given in 
figure 1.

\begin{table*}
\begin{center}
\caption{Photometric and spectroscopic 
parameters for the program stars.}
\begin{tabular}{lcccccccccccc}
\tableline\tableline
Parameters & LEID 34225 & LEID 39048 & LEID 61067 & LEID 32169 &\\
\tableline
RA & 13 27 53.7 & 13 26 3.9 & 13 26 50.7 & 13 27 33.2 \\
Declination & -47 24 43.3 & -47 26 54.1 & -47 37 1.0 & -47 23 47.9 \\
Visual Magnitude (V) & 13.0 & 12.8 & 12.5 & 13.3 \\
$B-V$ & 1.23 & 1.42 & 1.60 & 1.17\\
Metallicity([Fe/H]) & -1.0 & -0.65 & -1.0 & -1.0 \\ 
V$_{helio}$ & 235$\pm$0.5 & 238$\pm$0.5 & 230$\pm$0.5 & 230$\pm$0.5\\
Date of Observation & 2016 May 27 & 2016 May 3& 2016 May 15 & 2016 June 25\\
$T_{\rm eff}$\tablenotemark{a} & 4275$\pm$50 & 3965$\pm$50 & 
4040$\pm$50 & 4285$\pm$50 \\
log $g$\tablenotemark{a} & 1.30$\pm$0.1 & 0.95$\pm$0.1 & 0.85$\pm$0.1
& 1.35$\pm$0.1 \\
$\xi$\tablenotemark{a} & 1.6$\pm$0.1 & 1.6$\pm$0.1 & 1.6$\pm$0.1 & 1.6$\pm$0.1 \\  
$T_{\rm eff}$\tablenotemark{b} & 4265$\pm$50 & 3965$\pm$50 & 
4035$\pm$50 & 4285$\pm$50 \\
log $g$\tablenotemark{b} & 1.30$\pm$0.15 & 0.95$\pm$0.15 & 0.85$\pm$0.15 & 1.35$\pm$0.15 \\
$\xi$\tablenotemark{b} & 1.6$\pm$0.2 & 1.6$\pm$0.2 & 1.6$\pm$0.2 & 1.6$\pm$0.2\\   
$T_{\rm eff}$\tablenotemark{c} & 4266$\pm$50 & 3945$\pm$50 & 4010$\pm$50 & 4260$\pm$50 \\
log $g$\tablenotemark{c} & 1.35$\pm$0.1 & 1.0$\pm$0.1 & 0.95$\pm$0.1 & 1.4$\pm$0.1 \\
\tableline
\end{tabular}
\end{center}
\scriptsize
{$^{a}$ This work \\
$^{b}$ From \citet{johnson10}\\
$^{c}$ Photometric determinations from \citet{hema14}}
\end{table*}

\begin{figure}
\epsscale{0.80}
\plotone{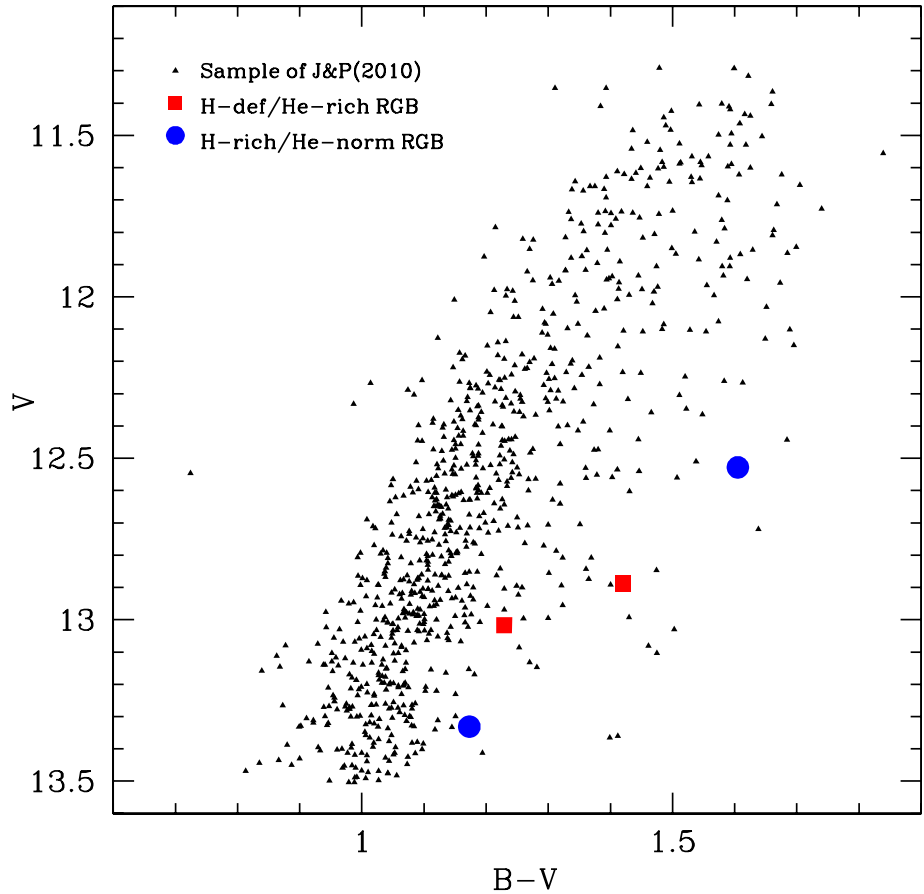}
\caption{Figure shows the Color-magnitude diagram for 
our sample $\omega$ Cen red giants along with the 
sample red giants from \citet{johnson10}. All the photometric
data are from \citet{johnson10}.
The filled red squares are the 
H-deficient/He-enhanced giants and the filled blue 
circles are the normal giants of our sample.
The filled black triangles are the sample red giants of
\citet{johnson10}. 
See figure for the keys.
\label{}}
\end{figure}

\clearpage
\begin{deluxetable}{ccccccccccccc}
\tabletypesize{\small}
\tablecolumns{7}
\tablewidth{0pc}
\tablecaption{The linelist used for the abundance analysis of the program stars. The table gives the wavelength (\AA), lower excitation potentail (eV), transition probabilities (log $gf$), and the measured equivalent widths/log\,$\epsilon$(E) for that line, for different species.}
\tablehead{
\colhead{Wavelength} & \colhead{$\chi$(eV)} & \colhead{log $gf$} & \colhead{LEID\,34225} & \colhead{LEID\,39048} & \colhead{LEID61067} & \colhead{LEID32169}\\
(\AA) & (eV) & & (m\AA)/log\,$\epsilon$(E)\tablenotemark{a} & (m\AA)/log\,$\epsilon$(E)\tablenotemark{a} & (m\AA)/log\,$\epsilon$(E)\tablenotemark{a} & (m\AA)/log\,$\epsilon$(E)\tablenotemark{a}}
\startdata
O\,{\sc i} $\lambda$6300.31 & 0.00 & $-$9.75 & 13/7.40 & 43/8.06 & 33/5.65 & 21/7.71\\ 
O\,{\sc i} $\lambda$6363.78 & 0.02 & $-$10.19 & \nodata  & 21/8.15 & \nodata  & 10/7.81\\
{\bf Mean (log\,$\epsilon$(O\,{\sc i}))} & \nodata & \nodata & 7.4 & 8.1$\pm$0.06 & 7.73 & 7.76$\pm$0.07 \\
Na\,{\sc i} $\lambda$4751.82\tablenotemark{b} & 2.10 & $-$2.08 & 26/5.73 & \nodata & 79/6.46 & 34/5.91 \\
Na\,{\sc i} $\lambda$5148.84\tablenotemark{b} & 2.10 & $-$2.04 & 39.7/5.93 & \nodata & 76/6.31 & 41/5.96 \\
Na\,{\sc i} $\lambda$6154.23 & 2.10 & $-$1.57 & 58/5.67 & 134/6.53 & 122/6.42 &  79/6.00\\
Na\,{\sc i} $\lambda$6160.75 & 2.10 & $-$1.27 & 70/5.55 & 149/6.47 & 142/6.44 & 98/5.97 \\
{\bf Mean (log\,$\epsilon$(Na\,{\sc i}))}& \nodata & \nodata & 5.72$\pm$0.16 & 6.50$\pm$0.04 & 6.40$\pm$0.07 & 5.96$\pm$0.04 \\ 
Mg\,{\sc i} $\lambda$4730.04\tablenotemark{b} & 4.34 & $-$2.39 & \nodata & 97/7.27 & 89/7.10 & \nodata\\
Mg\,{\sc i} $\lambda$5711.09\tablenotemark{b} & 4.34 & $-$1.73 & 105/6.70 & 145/7.35 & 121/6.86 & 115/6.88\\
Mg\,{\sc i} $\lambda$6318.72\tablenotemark{b} & 5.11 & $-$1.73 & 40/6.60 & 83/7.52 & 69/7.01 & 39/6.61 \\ 
Mg\,{\sc i} $\lambda$6319.24\tablenotemark{b} & 5.11 & $-$1.95 & 29/6.61 & 49/7.43 & 56/7.01 & 34/6.74 \\
Mg\,{\sc i} $\lambda$7657.61\tablenotemark{b} & 5.11 & $-$1.28 & 78/6.70 & 126/7.47 & 106/7.05 & 85/6.81 \\
{\bf Mean (log\,$\epsilon$(Mg\,{\sc i}))}& \nodata & \nodata & 6.65$\pm$0.05 & 7.41$\pm$0.1 & 7.00$\pm$0.1 & 6.80$\pm$0.14 \\
Al\,{\sc i} $\lambda$6696.03 & 3.14 & $-$1.57 & 100/6.37 &  119/6.37 & 110/6.30 & 79/6.05 \\
Al\,{\sc i} $\lambda$6698.67 & 3.14 & $-$1.89 & 89/6.49 &  103/6.42 & 90/6.30 & 66/6.14 \\
Al\,{\sc i} $\lambda$7835.31 & 4.02 & $-$0.64 & 93/6.66 & 99/6.56 & 73/6.23 & 59/6.15 \\
Al\,{\sc i} $\lambda$7836.13 & 4.02 & $-$0.49 & 91/6.47 & 115/6.65 & 108/6.59 & 75/6.24 \\
{\bf Mean (log\,$\epsilon$(Al\,{\sc i}))}& \nodata & \nodata & 6.50$\pm$0.12 & 6.50$\pm$0.13 & 6.35$\pm$0.16 & 6.15$\pm$0.08\\
Si\,{\sc i} $\lambda$5684.48\tablenotemark{b} & 4.95 & $-$1.65 & 58/6.91 & 71/7.45 & 59/7.01 & 48/6.76 \\
Si\,{\sc i} $\lambda$5690.42\tablenotemark{b} & 4.93 & $-$1.87 & 47/6.92 & 50/7.22 & 67/7.35 & 50/6.98\\
Si\,{\sc i} $\lambda$5701.10\tablenotemark{b} & 4.93 & $-$2.05 & 48/7.11 & 40/7.19 & 41/7.03 & 31/6.79\\
Si\,{\sc i} $\lambda$5772.15\tablenotemark{b} & 5.08 & $-$1.75 & 57/7.16 & 62/7.54 & 58/7.28 & 46/6.97\\
Si\,{\sc i} $\lambda$5793.07\tablenotemark{b} & 4.93 & $-$2.06 & 41/6.97 & 56/7.52 & 54/7.29 & \nodata \\
Si\,{\sc i} $\lambda$6155.13 & 5.62 & $-$0.78 & \nodata & 75/7.20 &  81/7.08 & 66/6.68\\
Si\,{\sc i} $\lambda$6237.32 & 5.61 & $-$1.28 & \nodata & 80/7.41 &  74/7.07 & 69/6.85\\
{\bf Mean (log\,$\epsilon$(Si\,{\sc i}))}& \nodata & \nodata & 7.00$\pm$0.11 & 7.36$\pm$0.15 & 7.16$\pm$0.14 & 6.84$\pm$0.12\\
Ca\,{\sc i} $\lambda$5260.39\tablenotemark{b} & 2.52 & $-$1.90 & 50/5.59 & \nodata & 78/5.82 & 57/5.73\\
Ca\,{\sc i} $\lambda$5867.56\tablenotemark{b} & 2.93 & $-$0.80 & 60/5.17 & \nodata & 89/5.42 & 89/5.68\\
Ca\,{\sc i} $\lambda$6161.29 & 2.52 & $-$1.28 & 83/5.20 & 143/5.85 & 140/5.89 & 98/5.47\\
Ca\,{\sc i} $\lambda$6162.18 & 1.90 & $-$0.07 & \nodata & 345/5.88 & 295/5.58 & 233/5.41\\
Ca\,{\sc i} $\lambda$6166.44 & 2.52 & $-$1.11 & 96/5.31 & 158/6.00 & 135/5.69 & 106/5.49\\
Ca\,{\sc i} $\lambda$6169.04 & 2.52 & $-$0.69 & 122/5.43 & 168/5.80 & 146/5.53 & 125/5.47\\
Ca\,{\sc i} $\lambda$6169.56 & 2.53 & $-$0.27 & 135/5.39 & 191/5.85 & 171/5.66 & 142/5.50\\
Ca\,{\sc i} $\lambda$6455.60\tablenotemark{b} & 2.52 & $-$1.34 &  90/5.62 & 137/6.01 & 111/5.66 &  95/5.70\\
Ca\,{\sc i} $\lambda$6471.66\tablenotemark{b} & 2.53 & $-$0.59 & 126/5.50 & 172/5.84 & 151/5.59 & 136/5.67\\
Ca\,{\sc i} $\lambda$6499.65\tablenotemark{b} & 2.52 & $-$0.59 & 124/5.45 & 176/5.88 & \nodata & 134/5.61\\
{\bf Mean (log\,$\epsilon$(Ca\,{\sc i}))}& \nodata & \nodata & 5.40$\pm$0.16 & 5.9$\pm$0.08 & 5.65$\pm$0.14 & 5.57$\pm$0.12 \\
Sc\,{\sc i} $\lambda$4743.82\tablenotemark{b} & 1.45 & 0.07 & \nodata & \nodata & 91/2.34 & \nodata\\
Sc\,{\sc i} $\lambda$5081.56\tablenotemark{b} & 1.45 & 0.06 & \nodata & \nodata & 79/2.10 & \nodata\\
Sc\,{\sc i} $\lambda$5484.63\tablenotemark{b} & 1.85 & 0.08 & 20/2.47 & \nodata & 32/2.38 & \nodata\\
Sc\,{\sc i} $\lambda$5671.83\tablenotemark{b} & 1.45 & 0.64  & 69/2.22 & \nodata &  101/2.32 & 59/2.08\\
Sc\,{\sc i} $\lambda$6210.67 & 0.00 & $-$1.53 & 58/2.08 & \nodata & 100/2.05 & 69/2.27 \\
Sc\,{\sc i} $\lambda$6305.66 & 0.02 & $-$1.30 & 85/2.17 & \nodata & 138/2.36 & 67/1.97 \\
{\bf Mean (log\,$\epsilon$(Sc\,{\sc i}))}& \nodata & \nodata & 2.24$\pm$0.16 & \nodata & 2.26$\pm$0.14 & 2.10$\pm$0.15 \\
Sc\,{\sc ii} $\lambda$5357.20\tablenotemark{b} & 1.51 & $-$2.21 & 19/2.36 & \nodata & 17/2.21 & 18/2.38\\
Sc\,{\sc ii} $\lambda$5552.23\tablenotemark{b} & 1.46 & $-$2.28 & \nodata & \nodata & 32/2.57 & 18/2.36 \\
Sc\,{\sc ii} $\lambda$5684.21\tablenotemark{b} & 1.51 & $-$1.07 & 74/2.29 & 69/2.28 & 72/2.16 & 56/2.00 \\
Sc\,{\sc ii} $\lambda$6245.62 & 1.51 & $-$0.98 & 80/2.27 & 95/2.60 & 86/2.25 & 60/1.97\\
Sc\,{\sc ii} $\lambda$6300.75 & 1.51 & $-$1.84 & 29/2.19 & 49/2.67 &  35/2.23 &  31/2.28 \\
Sc\,{\sc ii} $\lambda$6320.84 & 1.50 & $-$1.77 &  30/2.14 & 45/2.51 & 38/2.19 & \nodata \\
Sc\,{\sc ii} $\lambda$6604.58 & 1.36 & $-$1.48 & \nodata & 96/2.85 & 79/2.39 & \nodata \\
{\bf Mean (log\,$\epsilon$(Sc\,{\sc ii}))}& \nodata & \nodata & 2.25$\pm$0.09 & 2.60$\pm$0.2 & 2.28$\pm$0.14 & 2.20$\pm$0.2 \\
Ti\,{\sc i} $\lambda$5219.70\tablenotemark{b}  & 0.02 & $-$2.29 &  140/4.49 & 200/4.93 & 185/4.84 & \nodata \\
Ti\,{\sc i} $\lambda$5295.77\tablenotemark{b}  & 1.07 & $-$1.63 & 93/4.44 & 141/4.82 & 120/4.47 & 78/4.18 \\
Ti\,{\sc i} $\lambda$5490.15\tablenotemark{b}  & 1.46 & $-$0.93 &  99/4.38 & \nodata & 129/4.48 & 87/4.17\\
Ti\,{\sc i} $\lambda$5702.66\tablenotemark{b}  & 2.30 & $-$0.44 &\nodata & 99/4.54 & 67/4.10 & 31/3.86 \\
Ti\,{\sc i} $\lambda$5716.44\tablenotemark{b}  & 2.30 & $-$0.72 & 34/4.20 & 86/4.59 & 70/4.44 & 30/4.15\\
Ti\,{\sc i} $\lambda$6092.79\tablenotemark{b}  & 1.89 & $-$1.32 & \nodata & \nodata & 56/4.18 & 25/4.06\\
Ti\,{\sc i} $\lambda$6146.23  & 1.87 & $-$1.51 &  \nodata & \nodata & 70/4.17 & 22/3.78\\
Ti\,{\sc i} $\lambda$6258.11  & 1.44 & $-$0.38 &  \nodata & 197/4.75 & 161/4.20  & 105/3.74\\
Ti\,{\sc i} $\lambda$6261.10  & 1.43 & $-$0.49 &  \nodata & 210/5.05 & 169/4.46 & 104/3.82\\

Ti\,{\sc i} $\lambda$6303.76  & 1.44 & $-$1.69 &  58/4.19 & 145/5.03 & 95/4.27 & 56/4.19 \\
Ti\,{\sc i} $\lambda$6312.24  & 1.46 & $-$1.55 &  \nodata & 139/4.94 & 93/4.25 & 37/3.90 \\
Ti\,{\sc i} $\lambda$6336.11  & 1.44 & $-$1.69 &  \nodata & 126/4.85 & 81/4.25 & 30/3.94 \\
Ti\,{\sc i} $\lambda$6599.10\tablenotemark{b}  & 0.90 & $-$2.08 &  73/4.12 & 155/4.77 & 133/4.47 & 69/4.09 \\
Ti\,{\sc i} $\lambda$7357.73\tablenotemark{b}  & 1.44 & $-$1.12 & 98/4.22 & 166/4.65 & 169/4.77 & \nodata \\
Ti\,{\sc i} $\lambda$8675.37\tablenotemark{b}  & 1.07 & $-$1.67 & \nodata & \nodata & 162/4.29 & 100/4.09\\
Ti\,{\sc i} $\lambda$8682.98\tablenotemark{b}  & 1.05 & $-$1.94 & \nodata & \nodata & 149/4.37 & 66/3.94\\
Ti\,{\sc i} $\lambda$8734.71\tablenotemark{b}  & 1.05 & $-$2.38 & \nodata & \nodata & 138/4.66 & \nodata\\
{\bf Mean (log\,$\epsilon$(Ti\,{\sc i}))}& \nodata & \nodata & 4.30$\pm$0.14 & 4.80$\pm$0.17 & 4.40$\pm$0.2 & 4.0$\pm$0.16\\
Ti\,{\sc ii} $\lambda$4583.41\tablenotemark{b}  & 1.17 & $-$2.72 & 87/4.33 & \nodata & 99/4.52 & \nodata \\
Ti\,{\sc ii} $\lambda$4708.66\tablenotemark{b}  & 1.24 & $-$2.21 & 111/4.43 & \nodata & \nodata & 90/3.98 \\
Ti\,{\sc ii} $\lambda$5336.78\tablenotemark{b}  & 1.58 & $-$1.70 & 113/4.25 & \nodata & 121/4.34 & 107/4.16 \\
Ti\,{\sc ii} $\lambda$5418.77\tablenotemark{b}  & 1.58 & $-$1.99 &  95/4.16 & \nodata & 109/4.37 & 81/3.91\\
{\bf Mean (log\,$\epsilon$(Ti\,{\sc ii}))}& \nodata & \nodata & 4.29$\pm$0.11& \nodata & 4.40$\pm$0.1 & 4.02$\pm$0.13 \\
V\,{\sc i} $\lambda$6039.73\tablenotemark{b}  & 1.06 & $-$0.65 & \nodata & 125/3.26 & 113/3.13 & 54/2.73\\
V\,{\sc i} $\lambda$6081.44\tablenotemark{b}  & 1.05 & $-$0.58 & 66/2.81 & 137/3.39 & 124/3.23 & 63/2.78 \\
V\,{\sc i} $\lambda$6090.21\tablenotemark{b} & 1.08 & $-$0.06 & 74/2.44 & 149/3.15 & 132/2.90 & 87/2.65 \\
V\,{\sc i} $\lambda$6119.53\tablenotemark{b}  & 1.06 & $-$0.32 & 72/2.63 &  131/3.02 & 118/2.86 & 89/2.92\\
V\,{\sc i} $\lambda$6135.36\tablenotemark{b}  & 1.05 & $-$0.75 & 41/2.58 & 122/3.25 & 102/3.02 & 56/2.84\\
V\,{\sc i} $\lambda$6274.65\tablenotemark{b}  & 0.27 & $-$1.67 & \nodata & 139/3.18 & 129/3.08 & 73/2.84\\
V\,{\sc i} $\lambda$6285.16\tablenotemark{b}  & 0.28 & $-$1.51 & 81/2.77 & 154/3.33 & \nodata & 73/2.84\\
V\,{\sc i} $\lambda$6531.41\tablenotemark{b}  & 1.22 & $-$0.84 & 31/2.71 & 86/2.97 & 79/2.98 & 47/3.00\\
{\bf Mean (log\,$\epsilon$(V\,{\sc i}))}& \nodata & \nodata & 2.66$\pm$0.14 & 3.20$\pm$0.15 & 3.03$\pm$0.13 & 2.81$\pm$0.12 \\
Cr\,{\sc i} $\lambda$4708.02\tablenotemark{b}  & 3.17 & 0.11 & 68/4.50 & \nodata & \nodata & \nodata\\
Cr\,{\sc i} $\lambda$4801.05\tablenotemark{b}  & 3.12 & $-$0.13 & 61/4.51 & \nodata & 74/4.55 & 79/4.89\\
Cr\,{\sc i} $\lambda$4936.34\tablenotemark{b}  & 3.11 & $-$0.34 & 52/4.54 & \nodata & 70/4.64 & \nodata \\
Cr\,{\sc i} $\lambda$5272.01\tablenotemark{b}  & 3.45 & $-$0.42 & 40/4.82 & 63/4.93 &  53/4.83 & 23/4.45\\
Cr\,{\sc i} $\lambda$5287.20\tablenotemark{b}  & 3.44 & $-$0.90 & \nodata & 49/5.16 & 33/4.94 & \nodata\\
Cr\,{\sc i} $\lambda$5300.74\tablenotemark{b}  & 0.98 & $-$2.12 & 116/4.49 & 160/4.89 & 137/4.47 & 113/4.44 \\
Cr\,{\sc i} $\lambda$5304.18\tablenotemark{b}  & 3.46 & $-$0.68 & \nodata & 53/5.01 & 27/4.60 & 24/4.76\\
Cr\,{\sc i} $\lambda$5628.62\tablenotemark{b}  & 3.42 & $-$0.77 & \nodata & \nodata & \nodata & \nodata\\
Cr\,{\sc i} $\lambda$5781.16\tablenotemark{b}  & 3.01 & $-$2.15 & \nodata & \nodata & \nodata  & \nodata\\
Cr\,{\sc i} $\lambda$6882.48\tablenotemark{b}  & 3.44 & $-$0.38 & 42/4.70 & 75/4.94 & 51/4.61 & 30/4.48 \\
Cr\,{\sc i} $\lambda$6883.00\tablenotemark{b}  & 3.44 & $-$0.42 & 37/4.65 & \nodata & \nodata & 26/4.45 \\
{\bf Mean (log\,$\epsilon$(Cr\,{\sc i}))}&  \nodata & \nodata & 4.60$\pm$0.13 & 5.00$\pm$0.1 & 4.66$\pm$0.16 & 4.58$\pm$0.20 \\
Mn\,{\sc i} $\lambda$4671.69\tablenotemark{b}  & 2.89 & $-$1.66 & 32/4.45 & 65/4.83 & 34/4.26 & \nodata \\
Mn\,{\sc i} $\lambda$4709.71\tablenotemark{b}  & 2.89 & $-$0.34 &  91/4.32 & 125/4.89 & 111/4.58 & 92/4.35\\
Mn\,{\sc i} $\lambda$4739.11\tablenotemark{b}  & 2.94 & $-$0.49 & \nodata & 113/4.83 & 88/4.24 & 87/4.44 \\
Mn\,{\sc i} $\lambda$5004.89\tablenotemark{b}  & 2.92 & $-$1.64 & 31/4.40 & \nodata & 36/4.27 & \nodata \\
Mn\,{\sc i} $\lambda$5388.54\tablenotemark{b}  & 3.37 & $-$1.62 & \nodata & \nodata & 26/4.63 & \nodata \\
{\bf Mean (log\,$\epsilon$(Mn\,{\sc i}))}&  \nodata &  \nodata & 4.39$\pm$0.06 & 4.85$\pm$0.04 & 4.40$\pm$0.2 & 4.40$\pm$0.07 \\
Fe\,{\sc i} $\lambda$6151.61  & 2.17 & $-$3.28 & 85/6.12 & \nodata & 120/6.57 & \nodata \\
Fe\,{\sc i} $\lambda$6157.73  & 4.07 & $-$1.22 & 81/6.61 & \nodata &  \nodata & 80/6.61\\
Fe\,{\sc i} $\lambda$6165.36  & 4.14 & $-$1.46 & 49/6.34 & 76/6.89 & 49/6.26 & 46/6.31\\
Fe\,{\sc i} $\lambda$6173.34  & 2.22 & $-$2.89 & 114/6.34 & 146/6.84 &  129/6.40 & \nodata\\
Fe\,{\sc i} $\lambda$6180.20  & 2.73 & $-$2.66 & 96/6.48 & \nodata & \nodata & \nodata\\
Fe\,{\sc i} $\lambda$6187.99  & 3.94 & $-$1.67 & 50/6.29 & 80/6.88 & 68/6.53 & 55/6.42\\
Fe\,{\sc i} $\lambda$6200.32  & 2.61 & $-$2.41 & 118/6.50 & \nodata & \nodata & 108/6.33\\
Fe\,{\sc i} $\lambda$6219.28  & 2.20 & $-$2.42 & 144/6.41 & 186/6.99 & 161/6.49 &  139/6.35\\
Fe\,{\sc i} $\lambda$6226.74  & 3.88 & $-$2.19 & 31/6.38 & \nodata & 42/6.48 & 34/6.47\\
Fe\,{\sc i} $\lambda$6229.23  & 2.84 & $-$2.80 & 81/6.49 & \nodata & 86/6.40 & \nodata \\
Fe\,{\sc i} $\lambda$6232.64  & 3.65 & $-$1.23 & 111/6.64 & 130/7.00 & 113/6.55 & 105/6.53\\
Fe\,{\sc i} $\lambda$6246.32  & 3.60 & $-$0.85 & 139/6.71 & 135/6.64 &  126/6.36 & \nodata\\
Fe\,{\sc i} $\lambda$6252.56  & 2.40 & $-$1.67 & 172/6.40 & 190/6.63 &  \nodata & 180/6.57\\
Fe\,{\sc i} $\lambda$6265.14  & 2.18 & $-$2.56 & 154/6.69 & 177/6.96 & 153/6.44 & 138/6.43\\
Fe\,{\sc i} $\lambda$6270.22  & 2.86 & $-$2.60 & 99/6.66 & \nodata &  112/6.73 & 92/6.54\\
Fe\,{\sc i} $\lambda$6297.79  & 2.22 & $-$2.74 & \nodata & 161/6.92 & 133/6.31 & 127/6.44\\
Fe\,{\sc i} $\lambda$6301.50  & 3.65 & $-$0.72 & 132/6.52 & 152/6.88 & 141/6.57 & \nodata\\
Fe\,{\sc i} $\lambda$6302.49  & 3.69 & $-$1.11 & 99/6.33 & 135/7.04 &  117/6.56 & 100/6.35\\
Fe\,{\sc i} $\lambda$6311.50  & 2.83 & $-$3.17 & 48/6.27 & 77/6.69 & 72/6.48 & 55/6.40\\
Fe\,{\sc i} $\lambda$6315.81  & 4.07 & $-$1.69 & 57/6.62 & 76/7.00 &  56/6.51 & 65/6.77\\
Fe\,{\sc i} $\lambda$6322.69  & 2.59 & $-$2.41 & 123/6.54 & 145/6.86 & 134/6.55 & 113/6.37\\
Fe\,{\sc i} $\lambda$6335.33  & 2.20 & $-$2.17 & 150/6.24 & \nodata & 167/6.31 & \nodata\\
Fe\,{\sc i} $\lambda$6336.83  & 3.69 & $-$0.85 &  \nodata & 153/7.09 &  120/6.37 & 115/6.40\\
Fe\,{\sc i} $\lambda$6344.15  & 2.43 & $-$2.92 & \nodata & 126/6.76 & \nodata & \nodata\\
{\bf Mean (log\,$\epsilon$(Fe\,{\sc i}))}&  \nodata &  \nodata & 6.46$\pm$0.16 & 6.88$\pm$0.14 & 6.47$\pm$0.12 & 6.45$\pm$0.12 \\
Fe\,{\sc ii} $\lambda$6149.24  & 3.89 & $-$2.78 & 26/6.51 & \nodata & 20/6.49 & 21/6.39\\
Fe\,{\sc ii} $\lambda$6247.56  & 3.89 & $-$2.43 & 35/6.42 & \nodata &  33/6.55 & 38/6.54\\
{\bf Mean (log\,$\epsilon$(Fe\,{\sc ii}))}&  \nodata &  \nodata & 6.46$\pm$0.06 & \nodata & 6.52$\pm$0.04 & 6.46$\pm$0.11 \\
Co\,{\sc i} $\lambda$5212.69\tablenotemark{b}  & 3.51 & $-$0.11 & \nodata & 51/4.21 & 46/4.01 & \nodata\\
Co\,{\sc i} $\lambda$5280.63\tablenotemark{b}  & 3.63 & $-$0.03 & 36/3.98 & 60/4.48 & 33/3.83 & 20/3.65 \\
Co\,{\sc i} $\lambda$5301.04\tablenotemark{b}  & 1.71 & $-$2.00 & 66/3.92 & 88/4.20 & 92/4.20 & 69/4.01\\
Co\,{\sc i} $\lambda$5352.04\tablenotemark{b}  & 3.58 & 0.06 & 55/4.18 & 65/4.40 &  59/4.16 & 39/3.91\\
Co\,{\sc i} $\lambda$5647.23\tablenotemark{b}  & 2.28 & $-$1.56 & 41/3.78 & 68/4.18 &  66/4.03 & \nodata \\
Co\,{\sc i} $\lambda$6093.14\tablenotemark{b}  & 1.74 & $-$2.44 & 54/4.10 & 81/4.45  & 49/3.81 & 46/4.02 \\
Co\,{\sc i} $\lambda$6455.00\tablenotemark{b}  & 3.63 & $-$0.25 & 33/4.07 & 43/4.30 &  29/3.90 & 20/3.81\\
{\bf Mean (log\,$\epsilon$(Co\,{\sc i}))}& \nodata & \nodata & 4.01$\pm$0.14 & 4.32$\pm$0.13 & 4.00$\pm$0.16 & 3.88$\pm$0.15 \\
Ni\,{\sc i} $\lambda$5157.98\tablenotemark{b}  & 3.61 & $-$1.51 & 28/5.20 & \nodata &  42/5.42 & 28/5.23\\
Ni\,{\sc i} $\lambda$5537.11\tablenotemark{b}  & 3.85 & $-$2.20 & \nodata & \nodata & \nodata & \nodata\\
Ni\,{\sc i} $\lambda$6175.37  & 4.09 & $-$0.55 &  \nodata & 58/5.54 & 44/5.11 & 36/5.02\\
Ni\,{\sc i} $\lambda$6176.81  & 4.09 & $-$0.42 &  \nodata &  70/5.48 & 54/5.01 & 43/4.86\\
Ni\,{\sc i} $\lambda$6177.25  & 1.83 & $-$3.53 &  \nodata & \nodata & 47/4.95 & 40/5.07\\
Ni\,{\sc i} $\lambda$6186.71  & 4.11 & $-$0.96 &  \nodata & 47/5.75 & 30/5.25 & 19/5.04 \\
Ni\,{\sc i} $\lambda$6204.60\tablenotemark{b}  & 4.09 & $-$0.82 &  25/5.02 & 43/5.79 &  \nodata & 26/5.36\\
Ni\,{\sc i} $\lambda$6223.99\tablenotemark{b}  & 4.11 & $-$0.91 &  \nodata & 57/5.91 &  41/5.43 & 23/5.09\\
Ni\,{\sc i} $\lambda$6327.60  & 1.68 & $-$3.14 &  \nodata & 124/5.82 & 96/5.18 & 89/5.30\\
Ni\,{\sc i} $\lambda$6378.26  & 4.15 & $-$0.83 &  28/5.16 & 46/5.87 & 41/5.40 & 40/5.44\\
{\bf Mean (log\,$\epsilon$(Ni\,{\sc i}))}&\nodata & \nodata& 5.13$\pm$0.09 & 5.74$\pm$0.17 & 5.22$\pm$0.2 & 5.16$\pm$0.2 \\
La\,{\sc ii} $\lambda$6262.29 & 0.40 & $-$1.22 &  76/1.10  &  93/1.32 & 108/1.46 &  70/1.03\\
{\bf Mean (log\,$\epsilon$(La\,{\sc ii}))}&\nodata & \nodata& 1.10 & 1.32 & 1.46 & 1.03 \\
\enddata
\tablecomments
{$^{a}$ log\,$\epsilon$(E) is the abundance derived for that line.\\
$^{b}$ These lines are from \citet{ramirez11} and other lines are from 
\citet{johnson10}.} 
\end{deluxetable}

\clearpage
\begin{deluxetable}{cccc}
\tabletypesize{\small}
\tablecolumns{4}
\tablewidth{0pc}
\tablecaption{The list of Fe\,{\sc i} and Fe\,{\sc ii} lines from
\citet{ramirez11}. The Table gives the wavelength (\AA),
lower excitation potentail (eV), transition probabilities
(log gf), and the measured 
equivalent widths/log\,$\epsilon$(E) derived for that line
in the spectrum of program star LEID 34225.}
\tablehead{
\colhead{Wavelength} & \colhead{$\chi$} & \colhead{log $gf$} & \colhead{LEID\,34225} \\
(\AA) & (eV) & & (m\AA)/log\,$\epsilon$(E)\tablenotemark{a}} 
\startdata
Fe\,{\sc i} $\lambda$5295.31  & 4.42 & $-$1.59 & 25/6.39  \\
Fe\,{\sc i} $\lambda$5379.57  & 3.69 & $-$1.51 & 92/6.72 \\
Fe\,{\sc i} $\lambda$5386.33  & 4.15 & $-$1.67 & 31/6.26 \\
Fe\,{\sc i} $\lambda$5441.34  & 4.31 & $-$1.63 & 38/6.57 \\
Fe\,{\sc i} $\lambda$5705.46  & 4.30 & $-$1.35 & 37/6.26 \\
Fe\,{\sc i} $\lambda$5778.45 & 2.59 & $-$3.44 & 53/6.34 \\
Fe\,{\sc i} $\lambda$5793.91 & 4.22 & $-$1.62 & 55/6.74\\
Fe\,{\sc i} $\lambda$6003.01 & 3.88 & $-$1.06 & 95/6.50\\
Fe\,{\sc i} $\lambda$6027.05 & 4.08 & $-$1.09 & 78/6.43\\
Fe\,{\sc i} $\lambda$6056.00 & 4.73 & $-$0.40 & 73/6.54\\
Fe\,{\sc i} $\lambda$6079.01 & 4.65 & $-$1.02 & 38/6.38\\
Fe\,{\sc i} $\lambda$6093.64 & 4.61 & $-$1.30 & 40/6.64\\
Fe\,{\sc i} $\lambda$6096.66 & 3.98 & $-$1.81 & 64/6.76\\
Fe\,{\sc i} $\lambda$6151.62  & 2.17 & $-$3.28 & 85/6.12 \\
Fe\,{\sc i} $\lambda$6165.36  & 4.14 & $-$1.46 & 49/6.34\\
Fe\,{\sc i} $\lambda$6187.99  & 3.94 & $-$1.67 & 50/6.25 \\
Fe\,{\sc i} $\lambda$6240.65 & 2.22 & $-$3.29 & 102/6.52\\
Fe\,{\sc i} $\lambda$6270.22  & 2.86 & $-$2.60 & 99/6.60 \\
Fe\,{\sc i} $\lambda$6705.10 & 4.61 & $-$0.98 & 37/6.23 \\
Fe\,{\sc i} $\lambda$6713.74 & 4.79 & $-$1.40 & 24/6.61\\
Fe\,{\sc i} $\lambda$6726.67 & 4.61 & $-$1.03 & 49/6.47\\
Fe\,{\sc i} $\lambda$6810.26 & 4.61 & $-$0.98 & 49/6.47\\
Fe\,{\sc i} $\lambda$6828.59 & 4.64 & $-$0.82 & 51/6.37\\
Fe\,{\sc i} $\lambda$6842.69 & 4.64 & $-$1.22 & 43/6.62 \\
Fe\,{\sc i} $\lambda$6843.66 & 4.55 & $-$0.83 & 61/6.44\\
Fe\,{\sc i} $\lambda$7022.95 & 4.19 & $-$1.15 & 59/6.23\\
Fe\,{\sc i} $\lambda$7132.99 & 4.08 & $-$1.65 & 39/6.23\\
{\bf Mean (log\,$\epsilon$(Fe\,{\sc i}))}& \nodata & \nodata & 6.45$\pm$0.17\\
Fe\,{\sc ii} $\lambda$4576.33 & 2.84 & $-$2.95 & 77/6.24\\
Fe\,{\sc ii} $\lambda$4620.51 & 2.83 & $-$3.21 & 60/6.47\\
Fe\,{\sc ii} $\lambda$5234.62 & 3.22 & $-$2.18 & 79/6.35\\
Fe\,{\sc ii} $\lambda$5264.80 & 3.23 & $-$3.13 & 46/6.54\\
Fe\,{\sc ii} $\lambda$5425.26 & 3.20 & $-$3.22 & 42/6.52\\
Fe\,{\sc ii} $\lambda$6432.68 & 2.89 & $-$3.57 & 39/6.40\\
{\bf Mean (log\,$\epsilon$(Fe\,{\sc ii}))}& \nodata & \nodata  &6.49$\pm$0.1\\
\enddata
\tablecomments
{$^{a}$ log\,$\epsilon$(E) is the abundance derived for that line.} 
\end{deluxetable}

Figure 3 show the ($T_{\rm eff}$, log $g$)  plane 
for the program star LEID 34225, LEID 61067, LEID 32169.
For LEID 39048, no lines from ionized states were available. 
Hence, log $g$ value determined from photometric estimates 
from our previous studies \citet{hema14} were adopted. 
The log $g$ value derived by \citet{hema14} and 
\citet{johnson10} are in excellent agreement.
The uncertainties on the ($T_{\rm eff}$, log $g$)
for the program stars derived by \citet{johnson10}
are about $\pm$50 K and $\pm$0.15 (cgs), respectively,
and that derived photometrically by \citet{hema14} are about 
$\pm$100 K and $\pm$0.1 (cgs), respectively, which 
are in fair agreement.

\begin{figure}
\epsscale{0.8}
\plotone{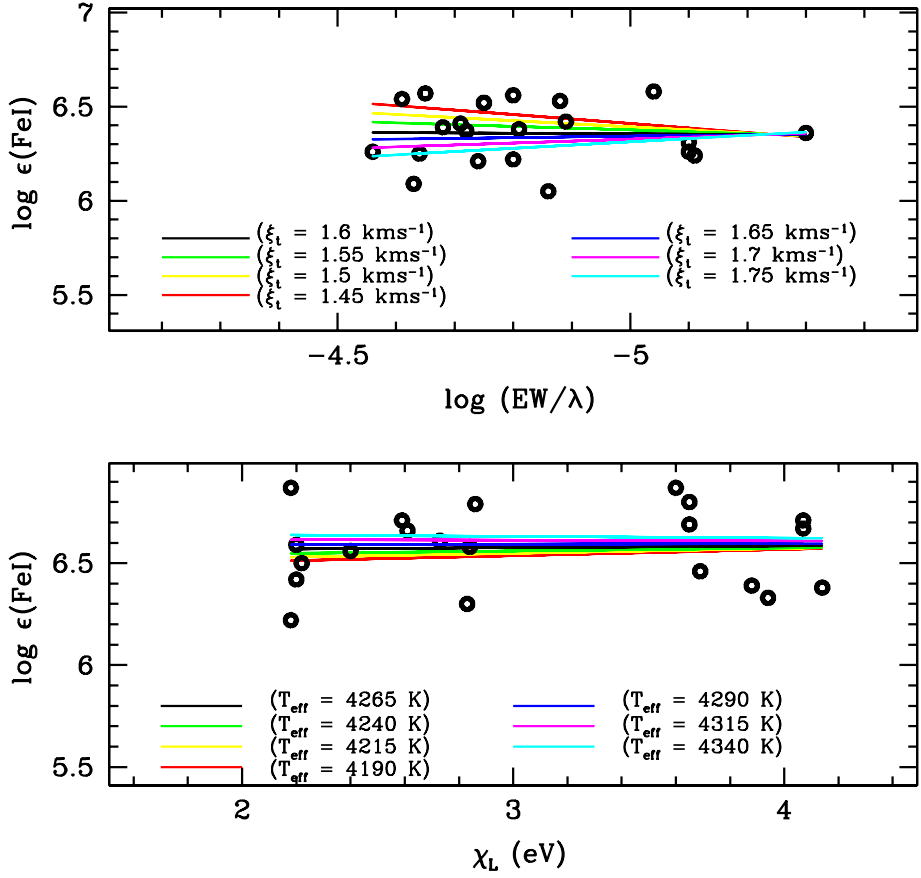}
\caption{Figure shows the estimation of uncertainties on
$T_{\rm eff}$ (lower panel) and $\xi_{t}$ (upper panel). 
The adopted stellar parameters for deriving 
$\Delta$$T_{\rm eff}$ and $\Delta$$\xi_{t}$ are,
($T_{\rm eff}$, log $g$, $\xi_{t}$):
(4265\,K, 1.3 cgs units, 1.6 kms$^{-1}$).  
The adopted uncertainties corresponding to the 
1$\sigma$ error on the abundance of the  
zero slope are $\Delta$$T_{\rm eff}$ = $\pm$50\,K
and $\Delta$$\xi_{t}$ = 0.1kms$^{-1}$.
See figure for the keys.
\label{}}
\end{figure}

\begin{figure}
\epsscale{0.8}
\plotone{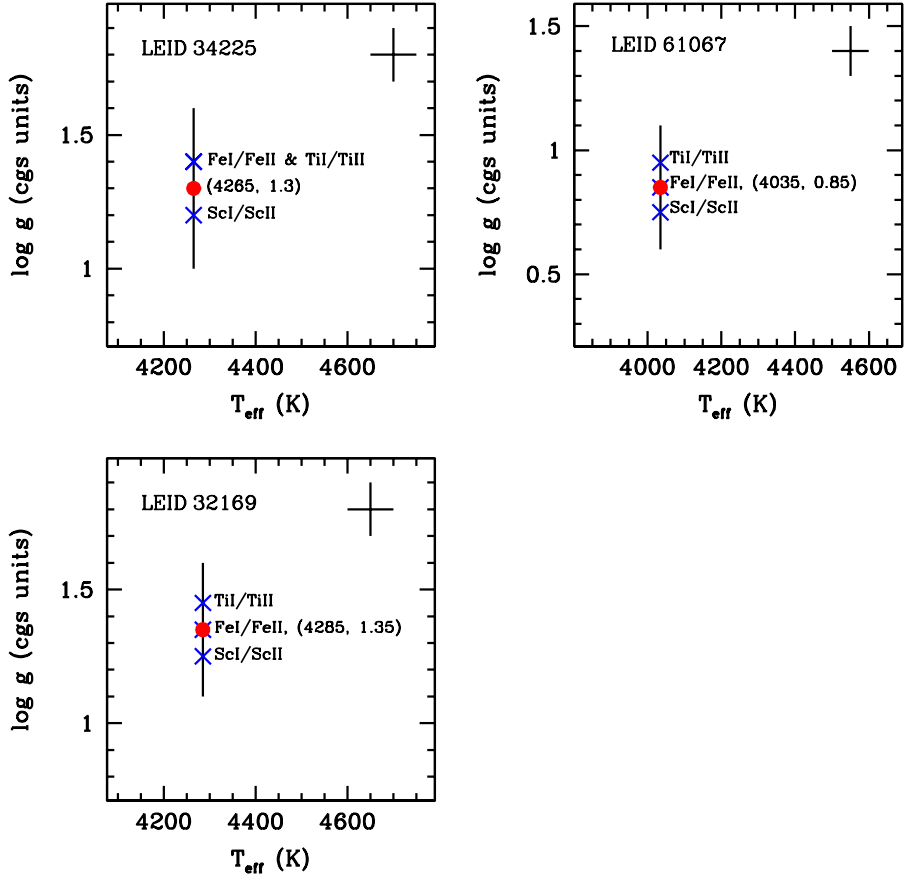}
\caption{Figure shows the plane of $T_{\rm eff}$ and log g for 
LEID 34225, LEID 61067, LEID 32169. The verticle line shows 
the excitation balance from Fe\,{\sc i} lines. 
The blue crosses on the verticle line
are the log $g$ values derived from the ionization balance of 
Fe\,{\sc i}/Fe\,{\sc ii}, Sc\,{\sc i}/Sc\,{\sc ii} and 
Ti\,{\sc i}/Ti\,{\sc ii}.
The mean/adopted ($T_{\rm eff}$, log $g$) is shown with a red dot. 
The error bars on the $T_{\rm eff}$ 
and log g  are shown in the top right corner. See figure for
the keys.
\label{}}
\end{figure}

For abundance analyses, the linelist of \citet{johnson10} 
was used. The elements for which the
lines are very few or none in \citet{johnson10} list, 
were adopted from \citet{ramirez11}. However, to cross examine, 
the stellar parameters and the Fe abundances were also derived 
from the Fe\,{\sc i} and Fe\,{\sc ii} lines of 
\citet{ramirez11} for one of the program stars 
i.e. LEID 34225. These parameters derived are in line 
with those determined from the
Fe\,{\sc i} and Fe\,{\sc ii} lines of \citet{johnson10}. 
The linelist used 
for the determination of the stellar parameters and the 
abundances for the program stars are given in Table 2. 
The Fe\,{\sc i} and Fe\,{\sc ii} lines of \citet{ramirez11} for
LEID 34225 are given in Table 3. Table 4 gives the 
abundance ratios ([E/Fe]) for different elements 
of the program stars. The adopted solar abundances 
from \citet{asplund09} is also given.
The typical errors on elemental abundances due to the 
uncertainties on the stellar parameters and the signal-to-noise 
are given in Table 5. 
The root-mean-square error due to these parameters
is given along with the standard deviation in the abundances due to 
line-to-line scatter, in the last two columns of Table 5.
An Mg\,{\sc i} line at 5711\AA\ in our program stars is also 
synthesized to support the Mg abundance derived from Mg\,{\sc i}
equivalent width analysis (see Figure 4).

\clearpage
\thispagestyle{empty}
\begin{deluxetable}{lccccccccccccc}
\tabletypesize{\small}
\tablecolumns{14}
\tablewidth{-40pc}
\tablecaption{The derived abundance ratios for the program stars. 
The adopted solar abundances from \citet{asplund09} are also given.}
\tablehead{
 \colhead{} & \colhead{Sun} &
\multicolumn{2}{c}{39048\tablenotemark{a}} & \colhead{} &
\multicolumn{2}{c}{61067\tablenotemark{c}} & \colhead{} &
\multicolumn{2}{c}{34225\tablenotemark{b}} & \colhead{} &
\multicolumn{2}{c}{32169\tablenotemark{c}} \\
\cline{3-4} \cline{6-7} \cline{9-10} \cline{12-13}
\colhead{Elements} &  \colhead{log$\epsilon$(E)} & \colhead{[E/Fe]}   & \colhead{n\tablenotemark{d}}  && \colhead{[E/Fe]} & \colhead{n\tablenotemark{d}}  && \colhead{[E/Fe]} & \colhead{n\tablenotemark{d}} && \colhead{[E/Fe]} & \colhead{n\tablenotemark{d}}}
\startdata
O\,{\sc i} & 8.69 & 0.03 & 2 && 0.14 & 1 && -0.05 & 1 && 0.11 & 2\\
Na\,{\sc i} & 6.24 & 0.88 & 2 && 1.17 & 4 && 0.52 & 4 && 0.77 & 4\\
Mg\,{\sc i} &  7.60 & 0.43 & 5 && 0.41 & 5  && 0.1 & 4 && 0.26 & 4 \\
Mg from MgH  & \nodata & 0.02 & \nodata && 0.26 & \nodata && -0.28 & \nodata && 0.25 & \nodata\\
Al\,{\sc i} & 6.45 & 0.67 & 4 && 0.92 & 4 && 1.09 & 4 && 0.75 & 4\\
Si\,{\sc i}  & 7.51 & 0.47 & 7 && 0.64 & 7 && 0.53 & 5 && 0.38 & 6 \\
Ca\,{\sc i} & 6.34 &  0.18 & 8 && 0.32 & 9 && 0.11 & 9 && 0.31 & 10 \\
Sc\,{\sc i} & 3.15 & \nodata & \nodata  && 0.12 & 6  && 0.13 & 4 && 0.0 & 3 \\
Sc\,{\sc ii} & \nodata & 0.06 & 5 && 0.14 & 7 && 0.14 & 5 && 0.1 & 5 \\
Ti\,{\sc i} & 4.95 & 0.48 & 11 && 0.46 & 17 && 0.39 & 7 && 0.1 & 14 \\
Ti\,{\sc ii} & \nodata & \nodata & \nodata && 0.46 & 3  && 0.39 & 4 && 0.1 & 3\\
V\,{\sc i} & 3.93 & -0.11 & 8 && 0.08 & 7 && -0.23 & 6 && -0.07 & 8\\
Cr\,{\sc i} & 5.64 & -0.02 & 5 && 0.02 & 7 && 0.0 & 7 && 0.01 & 6\\
Mn\,{\sc i} &  5.43 & 0.04 & 3 && -0.02 & 5 && 0.01 & 3 && 0.02 & 2\\
Fe\,{\sc i} & 7.50 & -0.62 & 16 && -1.01 & 19 && -1.05 & 21 && -1.05 & 16\\
Fe\,{\sc ii} &  \nodata & \nodata & \nodata && -1.05 & 2 && -1.05 & 2 && -1.06 & 2\\
Co\,{\sc i} &  4.99 & -0.06 & 7 && 0.02 & 8 && 0.05 & 6 && -0.04 & 5\\
Ni\,{\sc i} &  6.22 & 0.14 & 7 &&  -0.01 &  8 && -0.06 & 3 && -0.02 & 9\\
La\,{\sc ii} & 1.10 & 0.62 & 1 && 1.21 & 1 && 1.04 & 1 && 1.0 & 1\\
\enddata
\tablecomments
{$^{a}$ First group H-deficient star. \\
$^{b}$ Third group H-deficient star.\\
$^{c}$ First group normal stars.\\
$^{d}$ n is the number of lines used in the analysis.}
\end{deluxetable}

\begin{table*}
\begin{center}
\caption{Errors due to the uncertainties on the stellar parameters of
LEID 34225, defined by $\Delta$(log $\epsilon_{i}$) = log $\epsilon_{i}$ (perturbed) - log $\epsilon_{i}$ (adopted). The adopted parameters are
$T_{\rm eff}$ = 4265K, log $g$ = 1.3 (cgs), and $\xi_{turb}$= 1.6 km\,s$^{-1}$. }
\begin{tabular}{cccccccccccc}
\tableline\tableline
Species & $T_{\rm eff}$=$-$50 & log $g$=$-$0.1 & $\xi_{turb}$=$-$0.1 & Error$_{S/N}$ & RMS\tablenotemark{a} & SD\tablenotemark{b}\\
& [K] & [cgs] & km\,s$^{-1}$ &  & & \\
\tableline
O\,{\sc i} &  -0.04 & -0.01 & 0.03 & 0.08 & 0.09 & \nodata \\
Na\,{\sc i} & -0.06 & -0.05 & 0.01 & 0.09 & 0.11 & 0.16 \\
Mg\,{\sc i} & -0.02 & -0.005 & 0.02 & 0.1 & 0.10 & 0.05\\
Mg(MgH)\tablenotemark{c} & \nodata & \nodata & \nodata & 0.05 & 0.05 & \nodata\\ 
Al\,{\sc i} & -0.04 & 0.0 & 0.025 & 0.08 & 0.10 & 0.12\\
Si\,{\sc i} & 0.03 & -0.025 & 0.02 & 0.09 & 0.10 & 0.11\\
Ca\,{\sc i} & -0.08 & -0.005 & 0.04 & 0.07 &  0.11 & 0.16\\
Sc\,{\sc i} & -0.01 & -0.04 & 0.04 & 0.09  & 0.10 & 0.16 \\
Sc\,{\sc ii} & -0.01 & -0.045 & 0.02 & 0.08 & 0.09 & 0.09\\
Ti\,{\sc i} & -0.1 & -0.005 & 0.035 & 0.1 & 0.13 & 0.14\\
Ti\,{\sc ii} & 0.01 & -0.04 & 0.07 & 0.08 & 0.11 & 0.11\\
V\,{\sc i} & -0.12 & -0.01 & 0.01 & 0.09 & 0.14 & 0.14\\
Cr\,{\sc i} & -0.07 &  0.0 & 0.02 & 0.09 & 0.11 & 0.13\\
Mn\,{\sc i} & -0.07 & -0.005 & 0.015 & 0.1 & 0.11 & 0.06\\
Fe\,{\sc i} & -0.04 & -0.025 & 0.045 & 0.1 & 0.12 & 0.16 \\
Fe\,{\sc ii} 0.08 & -0.055 & 0.015 & 0.08 & 0.13 & 0.06 \\
Co\,{\sc i} & -0.01 & -0.025 & 0.015 & 0.09 & 0.10 & 0.14\\
Ni\,{\sc i} & -0.06 & -0.02 & 0.015 & 0.1 & 0.11 & 0.1\\
La\,{\sc i} & -0.02 & -0.06 & 0.1 & 0.09 & 0.14 & \nodata\\
\tableline
\end{tabular}
\end{center}
{$^{a}$ Root-mean-square of the error  on $T_{\rm eff}$,
log $g$, $\xi_{t}$ and the error on Signal-to-noise ratio of the spectrum.}\\
{$^{b}$ SD, the standard deviation on the abundance due to the line-to-line scatter.}\\
{$^{c}$ The error on synthesis due to the error in signal-to-noise
is given. Errors due to uncertainties on the stellar parameters 
for MgH band is discussed in Table 6.} 
\end{table*}

\begin{figure}
\epsscale{1.0}
\plotone{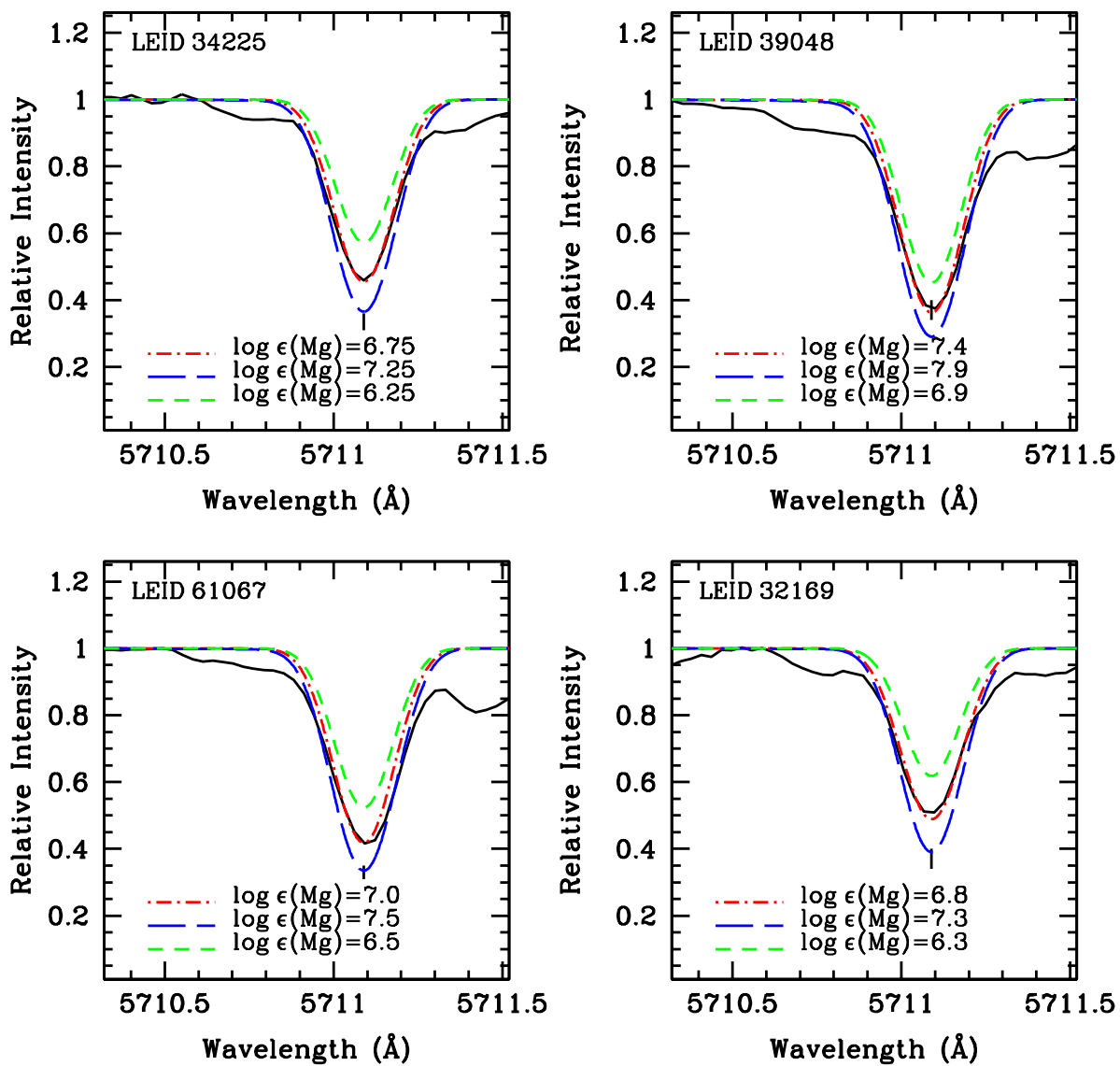}
\caption{Figure show the synthesis of Mg\,{\sc i} line
at 5711\AA\ for the program stars.  The synthesis is 
shown for the best fit Mg abundance 
and also for other two Mg abundances for comparison. 
See figure for the keys.
\label{}}
\end{figure}

\begin{table*}
\begin{center}
\caption{The derived Mg abundances from Mg\,{\sc i} lines and the
MgH band for the adopted and for the 
uncertainties on the stellar parameters. The adopted stellar 
parameters and their corresponding derived Mg abundances are 
given in boldface.}
\begin{tabular}{cccccccccccc}
\tableline\tableline
Stars & $T_{\rm eff}$ &  log $g$ & [Mg/Fe] from Mg\,{\sc i} & [Mg/Fe] from MgH \\
\tableline 
{\bf LEID 39048} & {\bf 3965} & {\bf 0.95} & {\bf $+$0.42} & {\bf $+$0.02} \\
& 4015 & 0.95 & $+$0.41 & $+$0.11 \\
& 3915 & 0.95 & $+$0.44 & $-$0.06 \\
& 3965 & 1.05 & $+$0.44 & $-$0.08 \\
& 3965 & 0.85 & $+$0.42 & $+$0.02 \\
{\bf LEID 34225} & {\bf 4265} & {\bf 1.30} & {\bf $+$0.10} & {\bf $-$0.30} \\
& 4315 & 1.30 & $+$0.10& $-$0.25 \\
& 4215 & 1.30 & $+$0.06 & $-$0.44 \\
& 4265 & 1.40 & $+$0.04 & $-$0.26 \\
& 4265 & 1.20 & $+$0.06 & $-$0.26 \\
\tableline
\end{tabular}
\end{center}
\end{table*}

\section{MgH band and the Spectrum syntheses}

In our previous study \citep{hema14}, the low-resolution 
optical spectra of the globular cluster 
$\omega$ Cen giants were analysed to identify the H-deficient stars
by examining the strengths of the Mg $b$ atomic lines and
the blue degraded (0,0) MgH band. Based on the strengths of 
these features, the observed program stars were devided 
into three groups, first: metal rich giants with strong Mg $b$ 
lines and also the MgH band; second: metal poor giants with 
no Mg $b$ line and no MgH band; third: metal rich giants with 
strong Mg $b$ lines and weak/no MgH band, in their observed spectra. 
\citet{hema14}'s analysis, that included comparison of stars'
spectra with similar stellar parameters and spectrum synthesis
of the MgH band for the star's adopted stellar parameters, 
resulted in identification of four mildly H-deficient stars: 
two from the first group (LEID 39048 and LEID 60073) and 
two from the third group (LEID 34225 and LEID 193804).

High resolution spectra of two stars, LEID 39048 -- a 
first group star, and LEID 34225 -- a third group star, 
including two comparison stars  
were obtained from SALT for confirming 
their H-deficiency. For example, SALT spectrum of 
LEID 34225 is superposed on the SALT spectrum of a 
comparison star, LEID 32169 (see Figure 5), note
that both these stars have similar stellar parameters 
i.e. effective temperature, surface gravity, and metallicity.
As discussed in \citet{hema14}, the strengths of MgH bands 
in the observed spectra of LEID 39048 and LEID 34225 
are weaker than that expected for their derived stellar 
parameters. To validate our continuum fitting in the MgH band
region, we measured the equivalent width for the Fe\,{\sc i} 
lines in this region and derived the abundances. 
The derived abundances are in excellent agreement with those 
derived from the other wavelength regions.

\begin{figure}
\begin{center}
\epsscale{0.9}
\plotone{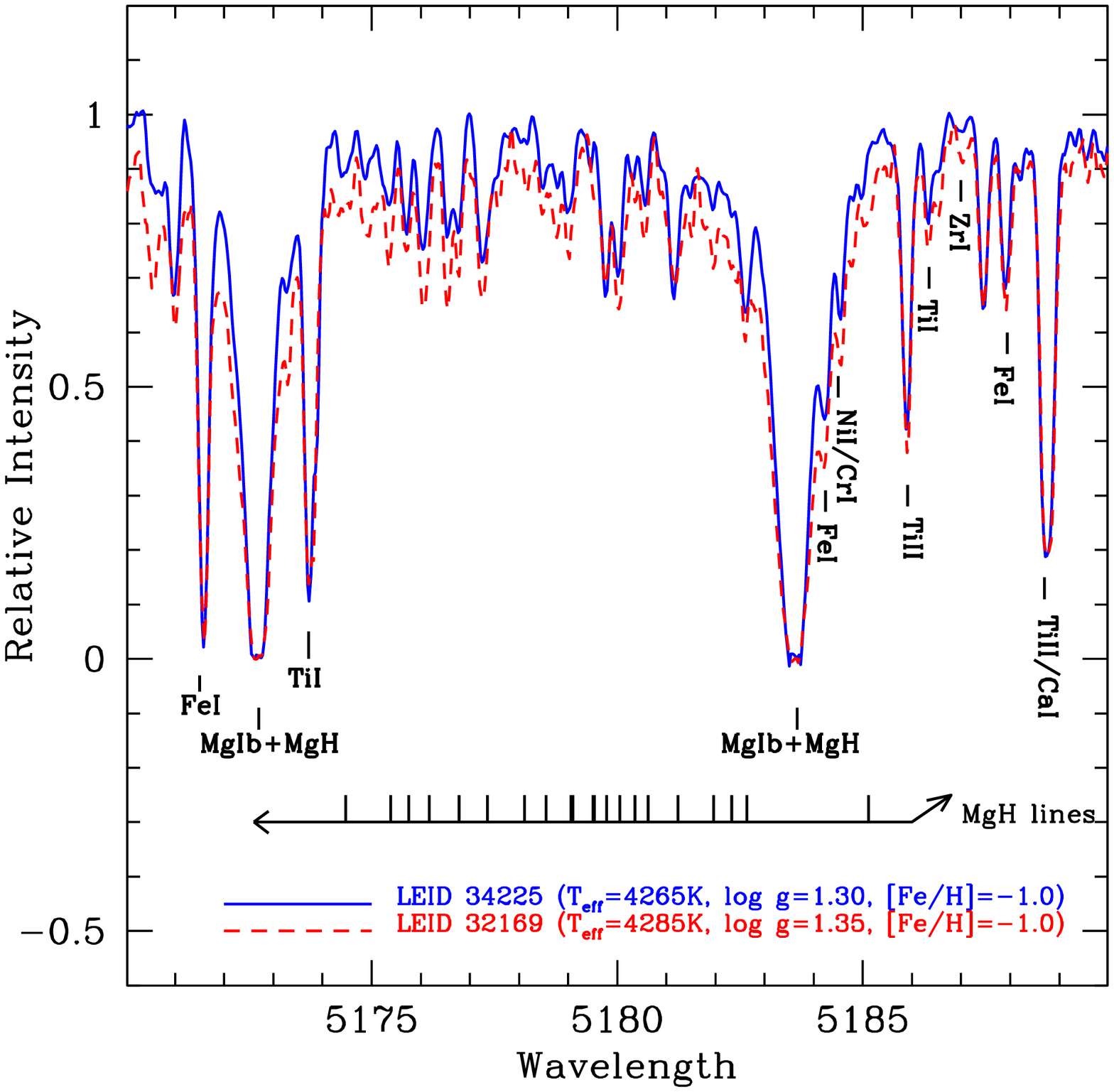}
\caption{Figure shows the strengths of (0,0) MgH band in the spectra
of first group stars: LEID 34225 (sample) and LEID 32169 (comparison). 
The key lines are marked.
\label{}}
\end{center}
\end{figure}

For their derived stellar parameters, the spectra of 
these stars were synthesized in the window 5170\AA\ to 5190\AA\
to examine the strengths of Mg $b$ lines and the MgH bands 
in their observed spectra carefully.
The spectrum synthesis for the program stars were carried out
following the procedure explained in \citet{hema14}.
Using the LTE spectrum synthesis code $synth$ in MOOG, 
combined with ATLAS9 \citep{kurucz98} plane parallel, 
line-blanketed LTE model atmospheres,
and the molecular and the atomic 
line lists both, validated by synthesizing the high-resolution 
spectrum of Arcturus provided by \citet{hinkle00}, were used to 
synthesize the spectra of our program stars. 
The Mg isotopic ratio, $^{24}$Mg:$^{25}$Mg:$^{26}$Mg;
82:09:09 was adopted for Arcturus from \citet{mcwilliam88}.
The synthesized spectrum was convolved with a 
Gaussian profile with a width that represents the 
instrumental broadening, as the effects due to
macroturbulence and the rotational velocity are very small
or negligible.

The spectra were synthesized in the wavelength window 
from 5170 to 5190\AA. 
The observed spectrum bluer to 5170\AA\ falls at 
the edge of the order. Our examination of the observed 
spectra show very strong saturated Mg $b$ lines at 
5167.3\AA\, 5172.68\AA\, and 5183.6\AA; The subordinate 
lines of MgH band are blended with these strong Mg $b$ lines.
Hence, the subordinate lines between the wavelength region 
5175\AA\ and 5176\AA\, where there are pure molecular lines, 
were given more weight. However, a fit to these MgH features 
gives an overall best fit to the whole range of MgH band 
spanning from about 5160\AA\ to 5190\AA.  The mean
of the isotopic ratios derived for red giants $\omega$ Cen by  
\citet{dacosta13}, which is about 
$^{24}$Mg:$^{25}$Mg:$^{26}$Mg;70:13:15 with the uncertainty 
on each value of about $\pm$4, was adopted for our program stars
that provides a fairly good fit through out the span of the 
MgH band. Note that, the Mg $b$ lines are very strong 
and are saturated in the spectra of our program stars, hence,
these lines are not used for estimating the Mg abundance
from Mg\,{\sc i} line or MgH band. The weaker atomic Mg\,{\sc i} 
lines are used for deriving the Mg\,{\sc i} abundance (those 
given in Table 2) and the Mg abundance from MgH band is derived 
using pure MgH molecular lines which are not blended with 
the strong Mg $b$ lines.

The spectra of the program stars, were synthesized 
using their derived stellar parameters and the elemental 
abundances as discussed in section 3. The syntheses of the 
spectra for the individual program stars are discussed below.

{\bf LEID 32169}: this is a first group star, which 
is relatively metal rich having strong Mg\,$b$ lines 
and a strong MgH band. This is a comparison for the 
first group mildly H-deficient stars and here for the 
sample star, LEID 34225. The spectrum of LEID 32169 
shows a well represented MgH band. Using its derived 
stellar parameters: ($T_{\rm eff}$, log $g$, $\xi_{t}$, 
[Fe/H]): (4285$\pm$50, 1.35$\pm$0.1, 1.6$\pm$0.1, -1.0), 
the MgH band is synthesized by varying the 
Mg abundance to obtain the best fit for the observed 
spectrum (see Figure 6). The MgH band synthesized for 
the log $\epsilon$(Mg) = 6.8 dex ([Mg/Fe]=$+$0.26) provides the best 
fit to the observed spectrum, and this is in excellent 
agreement with the Mg abundance derived from Mg\,{\sc i} 
lines. The Mg abundance derived from the MgH band is as 
expected for the star's metallicity and also that derived 
from the Mg\,{\sc i} lines.

\begin{figure}
\epsscale{0.8}
\plotone{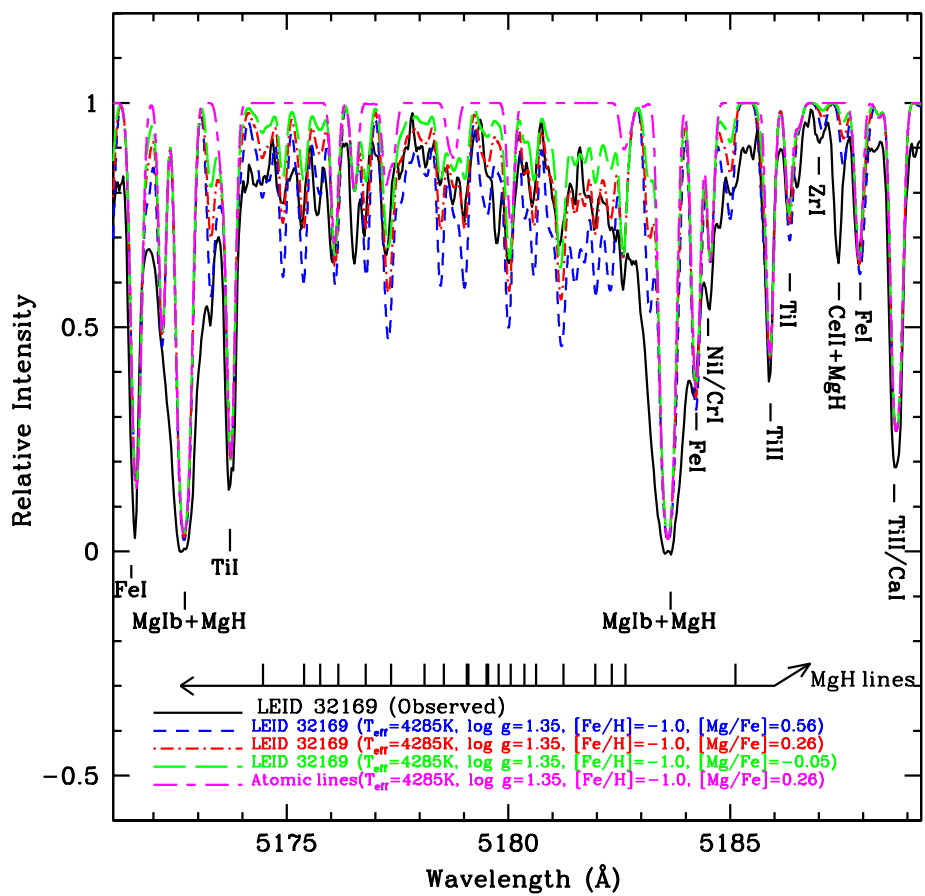}
\caption{Figure shows the superposition of the observed
and the synthesized spectra for the normal sample star LEID 32169.
The spectrum is synthesized for the star's derived
stellar parameters and the Mg abundance. The synthesis
is shown for the best  fit 
[Mg{\sc i}/Fe]=$+$0.26 with red dash-dotted line
and the syntheses for [Mg/Fe]=$+$0.56 is shown with blue 
short-dashed line, and for the [Mg/Fe]=$-$0.05 
is shown with green long-dashed line, for comparison.
The synthesis for pure atomic lines
is also shown with magenta short-long dashed line.
The key lines are marked. 
\label{}}
\end{figure}

{\bf LEID 61067}: This is a first group star, which 
is relatively metal rich having strong Mg $b$ lines 
and a strong MgH band. This is a comparison for the 
third group sample star LEID 39048.  
The observed spectrum shows the well represented
MgH band for its derived stellar parameters: ($T_{\rm eff}$, 
log $g$, $\xi_{t}$, [Fe/H]): (4035$\pm$50 K, 0.85$\pm$0.1, 
1.6$\pm$0.2, -1.0). The MgH band is synthesized 
by varying the Mg abundance. The spectrum synthesized for 
log $\epsilon$(Mg) = 6.85 dex ([Mg/Fe]=$+$0.25) provides the best fit to
the observed spectrum (see Figure 7). The derived Mg abundance from 
Mg\,{\sc i} lines is about 7.0$\pm$0.1 ([Mg/Fe] = $+$0.4). 
The Mg abundance 
required for obtaining the best fit for the observed 
spectrum is 6.85 dex and this is about 0.15 dex less than 
that derived from Mg\,{\sc i} lines, but are in fair agreement 
within the uncertainties on the derived abundances.

\begin{figure}
\epsscale{0.8}
\plotone{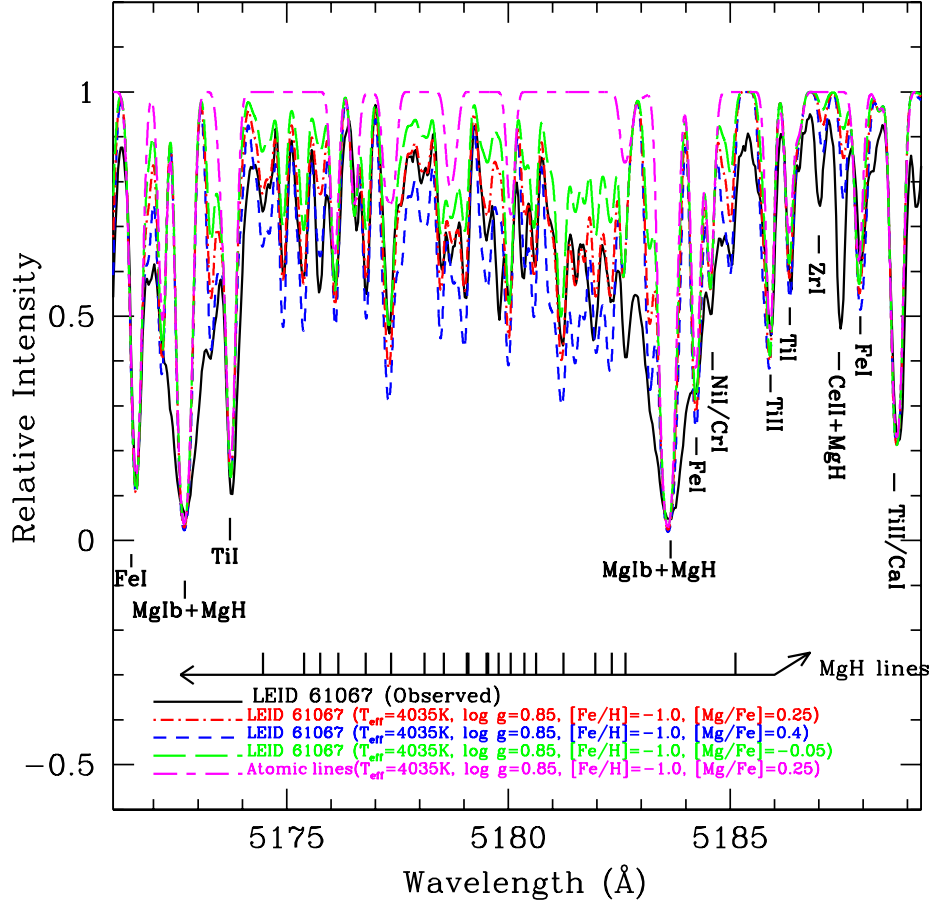}
\caption{Figure shows the superposition of the observed
and the synthesized spectra for the sample star LEID 61067.
The spectrum is synthesized for the star's derived
stellar parameters and the Mg abundance. The synthesis
is shown for [Mg/Fe]=$+$0.4 with blue
short-dashed line, the best fit [Mg/Fe]=$+$0.25 with
red dash-dotted line, and also for [Mg/Fe]=$-$0.05 with 
green long-dashed line for comparison. 
Note that the Mg abundance obtained 
from Mg\,{\sc i} and MgH are in fair agreement within the errors. 
The synthesis for pure atomic lines
is also shown with magenta short-long dashed line.
The key lines are marked. 
\label{}}
\end{figure}

{\bf LEID 39048}: This is the first group sample star, which is 
relatively metal rich having strong Mg $b$ lines and the 
MgH band.  From the previous study, that is the low resolution 
spectroscopic studies of $\omega$ Cen giants \citep{hema14}, 
this is one of the identified mildly H-deficient stars of 
our sample, having a very low Mg abundance as expected for
its metallicity as well as from the mean Mg abundance of the 
$\omega$ Cen giants. The Mg abundance was estimated from 
the MgH band for the star's derived stellar parameters.

In the present study we have used a high-resolution spectrum. 
The MgH band is synthesized for the star's derived stellar 
parameters: ($T_{\rm eff}$, log $g$, $\xi_{t}$, [Fe/H]): 
(3965$\pm$50 K, 0.95$\pm$0.1,
1.6$\pm$0.2, -0.65), and for the Mg abundance determined 
from Mg\,{\sc i} lines of 7.4 dex ([Mg/Fe]=$+$0.42). 
To obtain the best fit for the MgH band 
the Mg abundance has to be reduced  
by about 0.4 dex than that determined from Mg\,{\sc i} lines. 
This best fit value of the Mg abundance, derived from the 
MgH band, is beyond the uncertainty limit on the derived 
Mg abundance from Mg\,{\sc i} lines. Hence, this confirms our 
2014 results using a low resolution spectrum. Figure 8 shows 
the synthesis for LEID 39048 for the best fit 
Mg abundance and also for the derived Mg abundance from 
Mg\,{\sc i} lines, including a  different Mg abundance, which
do not provide the best fit to the observed MgH band.

 Spectrum of the program stars LEID 39048 
in the MgH band region was synthesized by 
changing the stellar parameters within the uncertainties 
that are discussed in section 3. Our aim was to explore
whether the difference in Mg abundance, from Mg\,{\sc i}
lines and from MgH band, could be accounted by making  
changes in the adopted stellar parameters within the
uncertainties. The synthesized MgH band for the uncertainties 
on effective temperature that are $+$50K and $-$50K, provides 
the best fit for the Mg abundance which is about 0.3 dex  
and 0.5 dex  lower than the Mg abundance from Mg\,{\sc i} 
lines, respectively (see Table 6). 
Similarly, the MgH band is also synthesized for 
the uncertainties on the
log $g$ value that are $\pm$0.1. For the log $g$ values $-$0.1
and $+$0.1 of the adopted log $g$, the synthesized MgH band provides 
the best fit for the Mg abundance which is about 0.4 
 and 0.5 dex lower than that derived from the 
Mg\,{\sc i} lines, respectively (see Table 6). 

Similar excercise was done with the uncertainties on the 
microturbulence; the strength of the synthesized MgH band 
show no appreciable difference with the change in microturbulence.

\begin{figure}
\epsscale{0.8}
\plotone{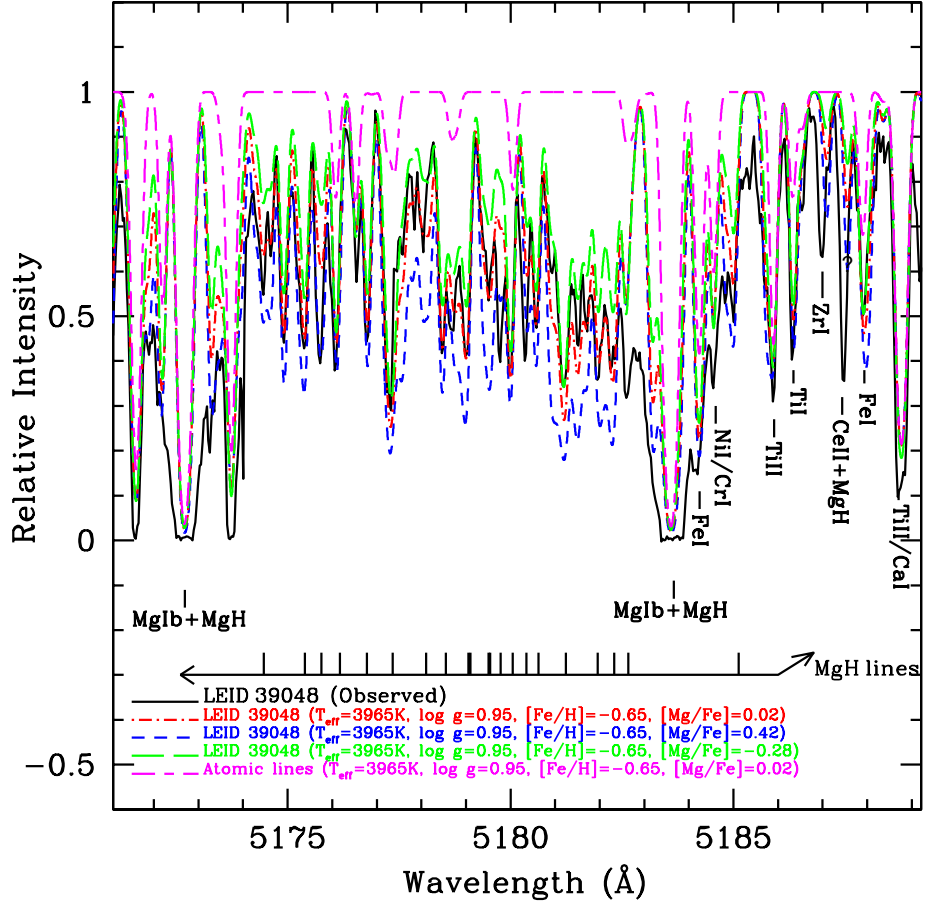}
\caption{Figure shows the superposition of the observed
and the synthesized spectra for the sample star LEID 39048.
The spectrum is synthesized for the star's derived
stellar parameters and the Mg abundance. The synthesis
is shown for [Mg/Fe]=$+$0.42 from Mg\,{\sc i} lines with blue 
short-dashed line, the best fit [Mg/Fe]=$+$0.02 with 
red dash-dotted line, and also for [Mg/Fe]=$-$0.28 with 
green long-dashed line, for comparison. 
The synthesis for pure atomic lines.
is also shown with magenta short-long dashed line.
The key lines are marked. 
\label{}}
\end{figure}

{\bf LEID 34225}: This is a third group sample star which 
is relatively metal rich having strong Mg $b$ lines 
and the MgH band. From our previous study \citep{hema14}, 
that is the low resolution spectroscopic 
studies of $\omega$ Cen giants, this is one of the mildly
H-deficient stars of our sample giving a very low Mg abundance 
than expected for its metallicity and also from the 
mean Mg abundance of the $\omega$ Cen giants.  
This low Mg abundance was derived by synthesizing 
the MgH band for the star's adopted stellar parameters 
using the observed low resolution spectrum.

A high-resolution spectrum is used in the present study.
The MgH band is synthesized for the star's derived 
stellar parameters: ($T_{\rm eff}$, log $g$, $\xi_{t}$, [Fe/H]): 
(4265$\pm$50 K, 1.30$\pm$0.1, 1.6$\pm$0.1, -1.0)  
and for the Mg abundance determined 
from Mg\,{\sc i} lines of 6.65$\pm$0.05 dex ([Mg/Fe]=0.1). 
To obtain the best fit to 
the observed high-resolution spectrum, the Mg abundance 
has to be reduced by about 0.4 dex than that derived 
from Mg\,{\sc i} lines. This best fit value of the 
Mg abundance is outside the uncertainty limits on 
the Mg abundance from Mg\,{\sc i} lines. Hence, this 
confirms our 2014 results. Figure 9 shows the synthesis 
of the MgH band for the best fit Mg abundance, for the derived
Mg abundance from Mg\,{\sc i} lines, and also for a lower Mg 
abundance than that provided the best fit.

The spectra of the MgH band region were synthesized 
by changing the stellar parameters within  the 
uncertainties that are discussed in section 3.
Our aim was to explore whether the difference
in Mg abundance, from Mg\,{\sc i} lines and from MgH 
band, could be accounted by making changes in the 
adopted stellar parameters within  the uncertainties.  
The synthesized MgH band for the uncertainties on
effective temperature that are $+$50K and $-$50K, provides
the best fit for the Mg abundance which is about 0.3 dex and 
0.5 dex lower than the Mg abundance from Mg\,{\sc i} lines, 
respectively (see Table 6).
Similarly, the MgH band is also synthesized for the 
uncertainty on the log $g$ value that are $\pm$0.1. 
For the log $g$ values, $-$0.1 and $+$0.1 of the 
adopted log $g$, the synthesized MgH band provides 
the best fit for the Mg abundance which is about 
0.3 dex  lower than that derived from the Mg\,{\sc i} 
lines, respectively (see Table 6).

Similar excercise was done with the uncertainties 
on the microturbulence; The strength of the synthesized 
MgH band show no appreciable difference with the change in
microturbulence.

\begin{figure}
\epsscale{0.8}
\plotone{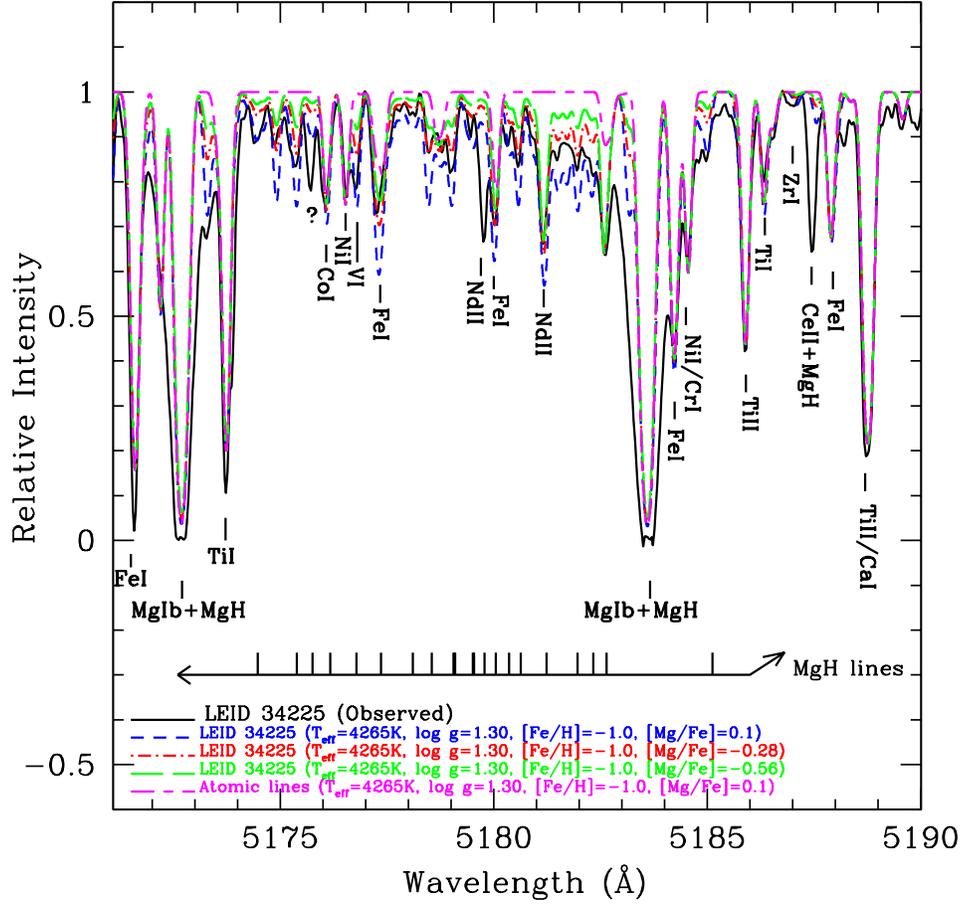}
\caption{Figure shows the superposition of the observed 
and the synthesized spectra for the sample star LEID 34225.
The spectrum is synthesized for the star's derived 
stellar parameters and the Mg abundance. The synthesis 
is shown for [Mg/Fe] = $+$0.1 from Mg\,{\sc i} lines
with blue short-dashed line, the best 
fit [Mg/Fe] of $-$0.28 with the red dash-dotted line,
and also for [Mg/Fe] of $-$0.56 with 
green long-dashed line for comparison.
The synthesis for pure atomic lines 
is also shown with magenta short-long dashed line. 
The key lines are marked. 
\label{}}
\end{figure}

\section{Discussion}

The aim of \citet{hema14} was to 
identify the H-deficient stars of RCB type, 
which show a severe H-deficiency. The features 
that directly indicate the H-deficiency  
are H-Balmer lines, CH-band, etc. H$\beta$ and CH-band 
were not covered in the observed low-resolution spectra, 
however, saturated H$\alpha$ line is present.
Hence, the (0,0) MgH band was used for the analysis. 
The region of (0,0) MgH band 
also includes the strong Mg\,$b$ lines, and these indicate 
the appropriate metallicity and the Mg abundance 
of the program stars. 
The four stars that were identified as H-poor, 
were confirmed by the spectrum synthesis. Though the accurate 
stellar parameters  and the metallicity were 
available for the program stars, the Mg abundances 
were not known. In this study, the stellar parameters 
and the elemental abundances, especially the Mg 
abundance from Mg\,{\sc i} lines, were rederived  
using the high-resolution spectra of the program stars 
obtained from SALT-HRS.

In the observed high-resolution spectra, initially we 
looked for the strengths of the H-Balmer line and the 
CH-band. But the H-Balmer lines in the observed spectra 
of  the program stars are strong and as 
expected for the stars' stellar parameters. The CH-band 
couldn't be detected in the spectra of these stars, due to  
poor signal at about 4300\AA. Hence, the analyses is 
based on the strength of the (0,0) MgH band in the observed 
SALT spectra.  
The two stars are mildly H-poor as expected. 
According to \citet{hema14}, the third group sample star 
LEID 34225 has strong Mg $b$ lines and weak/no
MgH band. The same traits are observed in the 
high-resolution spectrum of this star. And, the first group 
sample star LEID 39048 having strong Mg $b$ lines and strong MgH 
band, is also in line with its SALT high-resolution spectrum.
The observed MgH band of the program stars was analysed mainly
by spectrum synthesis. 

For LEID 32169, a normal (H-rich) comparison star, the best fit 
of the synthesized spectrum of the MgH band, for the 
star's adopted stellar parameters, to the observed 
spectrum is obtained  for the Mg abundance of 6.8 dex
(see Figure 6). The Mg abundance derived from the Mg\,{\sc i} 
lines and the MgH band are in excellent agreement, as expected. 
Similarly, for LEID 61067, a normal (H-rich) comparison star, the 
best fit of the synthesized spectrum of the MgH band, 
for the star's adopted stellar parameters, to the 
observed spectrum is obtained for the Mg abundance of 6.85 
(see Figure 7).  This Mg abundance from MgH band is about 
0.15 dex less than the derived Mg abundance from 
the Mg\,{\sc i} lines. This difference in abundance 
is within the uncertainties, which is about $\pm$0.1 dex
on the derived Mg abundance from Mg\,{\sc i} lines.

For LEID 39048, a candidate H-deficient star of our 
sample, the best fit of the synthesized spectrum 
of the MgH band, for the star's adopted stellar 
parameters, to the observed spectrum
is obtained for the Mg abundance of 7.0 dex ([Mg/Fe]=0.02 dex)
(see Figure 8). 
This Mg abundance is about 0.4 dex less than that 
derived from the Mg\,{\sc i} lines. This difference 
between the derived Mg abundance from Mg\,{\sc i} lines 
and that from MgH band is greater than the uncertainty
on the Mg abundances from Mg\,{\sc i} lines. 
The spectra were also synthesized by changing the 
stellar parameters within the uncertainties,  
the derived Mg abundance from Mg\,{\sc i} lines 
and that from the MgH band do not match even  within 
the uncertainties. The Mg abundance required to fit 
the observed spectrum for the adopted stellar parameters 
and the uncertainties on them, always require the 
Mg abundance that is lower by about or more than 0.3 dex 
than that derived from the Mg\,{\sc i} lines (see Table 6). This 
difference between the Mg abundance from Mg\,{\sc i} lines
and the MgH band is not acceptable, as the Mg abundance 
from Mg\,{\sc i} lines and that from
the MgH band are expected to be same  within the
uncertainties (as seen from the analysis of the
spectra of the normal comparison stars -- see above).

Similarly, for LEID 34225, another candidate  H-deficient star
of our sample, the best fit of the synthesized spectrum,
for the star's adopted stellar parameters, to the observed 
spectrum is obtained for the Mg abundance of 6.28 dex ([Mg/Fe]=$-$0.3)
(see Figure 9). This Mg abundance is about 0.4 dex less 
than that derived from the Mg\,{\sc i} lines. 
The spectra were also synthesized by 
changing the stellar parameters within the uncertainties.
The Mg abundance required to fit the observed 
spectrum for the adopted stellar parameters 
and the uncertainties on them, 
always require the Mg abundance that is lower by about
or more than 0.3 dex than that derived from the Mg\,{\sc i} lines 
(see Table 6).
This difference between the Mg abundance from Mg\,{\sc i} 
lines and the MgH band is not acceptable, as the Mg abundance 
from Mg\,{\sc i} lines and that from the MgH band
are expected to be same within the 
uncertainties.

The galactic globular cluster $\omega$ Cen is well known 
for hosting multiple stellar populations not only in
the red giant branch stars, but also in the main-sequence, 
and sub giant branches.  Among the red giant stars, about four 
distinct subpopulations are identified by \citet{calamida09}, 
viz. metal-poor; ([Fe/H] $\leq$ $-$1.49), metal-intermediate;
($-$1.49 $<$ [Fe/H] $\leq$ $-$0.93), metal-rich; 
($-$0.95 $<$ [Fe/H] $\leq$ $-$0.15), and solar metallicity; 
([Fe/H] $\approx$ 0). Among the subgiant branch stars, there 
are about three distinct subpopulations identified by 
\citet{villanova07}, viz. SGB-metal poor; ([Fe/H] $\sim$ $-$1.7),
SGB-Metal-intermediate; ($-$1.7 $<$ [Fe/H] $<$ $-$1.4), and  
SGB-a; ([Fe/H] $\sim$ $-$1.1).
Based on the metallcity, two main-sequences (MS) are identified
by \citet{piotto05}, viz. a red MS with ([Fe/H] $\sim$ $-$1.6), 
and a blue MS with ([Fe/H] $\sim$ $-$1.3).
There are also multiple stellar population studies in horizontal 
branch (HB) stars by \citet{tailo16}. They identify the 
HB stars as, metal-poor; ($-$2.25 $\leq$ [Fe/H] $\leq$ $-$1.4), 
metal-intermediate; ($-$1.4 $\leq$ [Fe/H] $\leq$ $-$1.1), 
and metal-rich; ([Fe/H] $\geq$ $-$1.1).

\begin{figure}
\epsscale{0.80}
\plotone{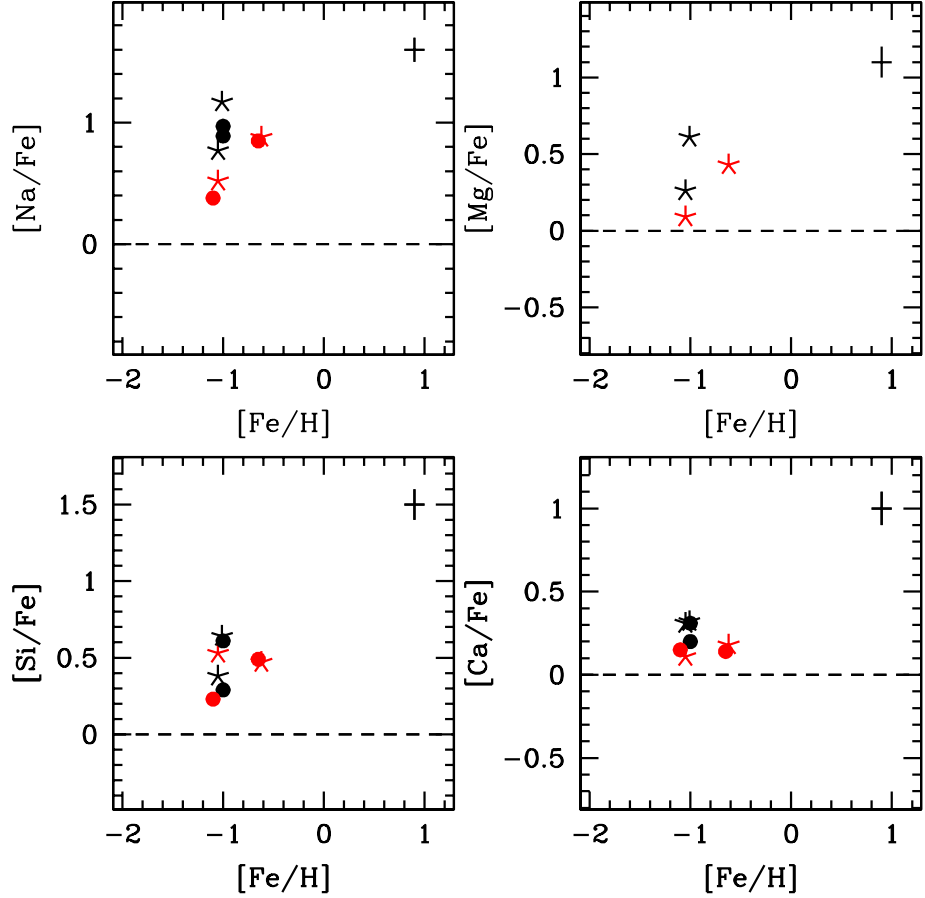}
\caption{The abundance ratios with respect to solar versus
metallicity for the program stars are shown. The red stars
are the mildly H-deficient stars and the black stars are the
normal comparison stars. The filled circles represents
the elemental abundances derived by \citet{johnson10} for
the program stars, the red circles the mildly
H-deficient/He-enhanced stars and black circles the normal
(H-rich) stars. The horizontal dotted lines shows
the solar scaled abundance values. The error bars on the
abundance ratios are shown with crossed thicklines in
the top right corner.
\label{}}
\end{figure}

According to these subpopulations, from the main-sequence,
through SGB, RGB, there is a metal rich group
having the metallicity ([Fe/H] $\sim$ $-$1.1$\pm$0.3), 
derived spectroscopically, which suggests 
that they are closely related. From the detailed  
studies of the main-sequence branches, it is revealed that, 
the bMS stars are helium enriched with an amount 
(0.35 $<$ Y $<$ 0.45) and are metal rich by about 0.3 dex than 
the majority red main-sequence stars which are He-normal (Y = 0.28). 
However, there are no helium enchancement studies reported  
for SGB stars, but the SGBs with metallicities
similar to the bMS stars are observed. And, these are identified 
as SGB-a group by \citet{villanova07}. \citet{villanova07} have 
compared their results of the abundance analyses of SGB-a stars 
with that of the bMS stars from \citet{piotto05}. The drived 
abundances, [C/Fe], [N/Fe], and [Ba/Fe], for bMS stars and 
SGB-a are very similar, ascertaining the connection between these
groups. \citet{pancino11} has studied metal rich subgiants
for determining their lithium abundance along with the $\alpha$-peak
elements, [Al/Fe] and [Ba/Fe].  These abundance ratios: [$\alpha$/Fe],
[Al/Fe] and [Ba/Fe] for their sample are in agreement with the
literature. They suspect that, all the H-burning processes, 
where He is produced, happens at temperatures where Li is destroyed. 
Therefore, He-rich stars should have a very low lithium content.  
In their sample, they have found a lithium abundance
which is lower than that expected for these metal-rich subgiants
(see \citet{pancino11}for details).
Hence, it is an indirect clue that, the metal-rich subgiant stars 
are He-rich.

\begin{figure}
\epsscale{0.80}
\plotone{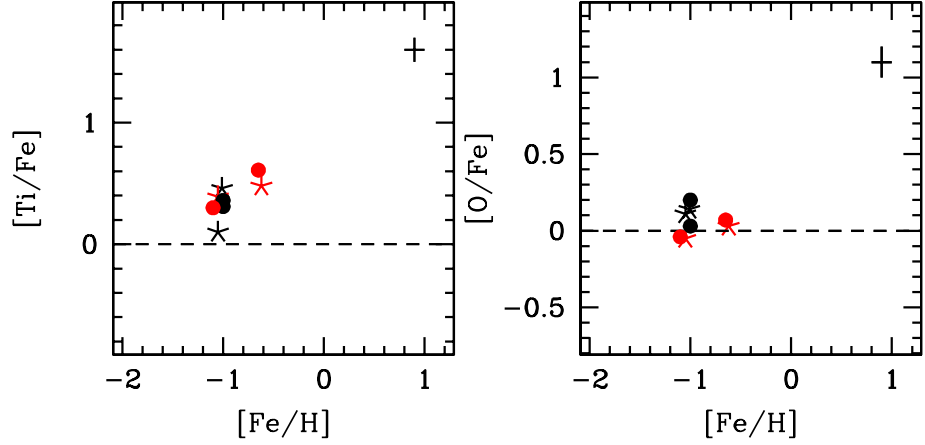}
\caption{The abundance ratios with respect to solar versus
metallicity for the program stars are shown. The red stars
are the mildly H-deficient stars and the black stars are the
normal comparison stars. The filled circles represents
the elemental abundances derived by \citet{johnson10} for
the program stars, the red circles the mildly
H-deficient/He-enhanced stars and black circles the normal
(H-rich) stars. The horizontal dotted lines shows
the solar scaled abundance values. The error bars on the
abundance ratios are shown with crossed thicklines in the
top right corner.
\label{}}
\end{figure}

In connection to evolution of the metal rich stars, 
our program stars provide further link to this 
evolutionary track.  Our program stars are 
metal rich with the metallicity ([Fe/H] $\geq$ $-$1.1). Using 
the high-resolution spectra, a detailed abundance analyses is
carried out. In order to check the similarities of our 
program stars with the metal rich stars of MS;bMS, and SGB;SGB-a,
we compared their elemental abundances.
Mainly the $\alpha$- and the Fe-peak elements are compared 
as these remain unaltered in the course of the evolution of 
these stars. 
The elemental abundances determined for the program stars 
agree within $\pm$0.2 dex with those derived by \citet{johnson10}.
This uncertainty arises as the spectra used are obtained from 
different instruments with different resolution.
The solar-scaled elemental abundances plotted vs. metallicity for 
the program stars are shown in Figure 10, 11 and 12. 
The abundance ratios from \citet{johnson10} are shown
with filled circles.
\citet{johnson10} have adopted the solar abundances from 
\citet{anders89} and we have adopted the solar abundances
from \citet{asplund09}; differences due to the adopted solar
abundances are taken into account.
The $\alpha$ processed elements, such as Mg, Si and Ca show
enhacement by about 0.2 to 0.7 dex which is in fair agreement 
with those derived by \citet{johnson10} for the metal-rich RGB 
stars of $\omega$ Cen (see Figure 10). [Na/Fe] shows enhancement 
of about 0.5-1.2 dex for the program stars, which is as expected for
the metal rich stars from \citet{johnson10} (see Figure 10). 
[Ti/Fe] shows an enhancement on an average of 0.35 dex for the 
program stars (see Figure 11). \citet{johnson10}
finds that for the metal rich RGBs [Ti/Fe] $\approx$ 0.3 dex. 
The Fe-peak elements such as, Cr, Mn, Co, and Ni do not show 
any enhancement (see Figure 12). \citet{johnson10} has given the 
abundances of Ni for the RGBs  which is in good agreement with 
that of our program stars.

\begin{figure}
\epsscale{0.80}
\plotone{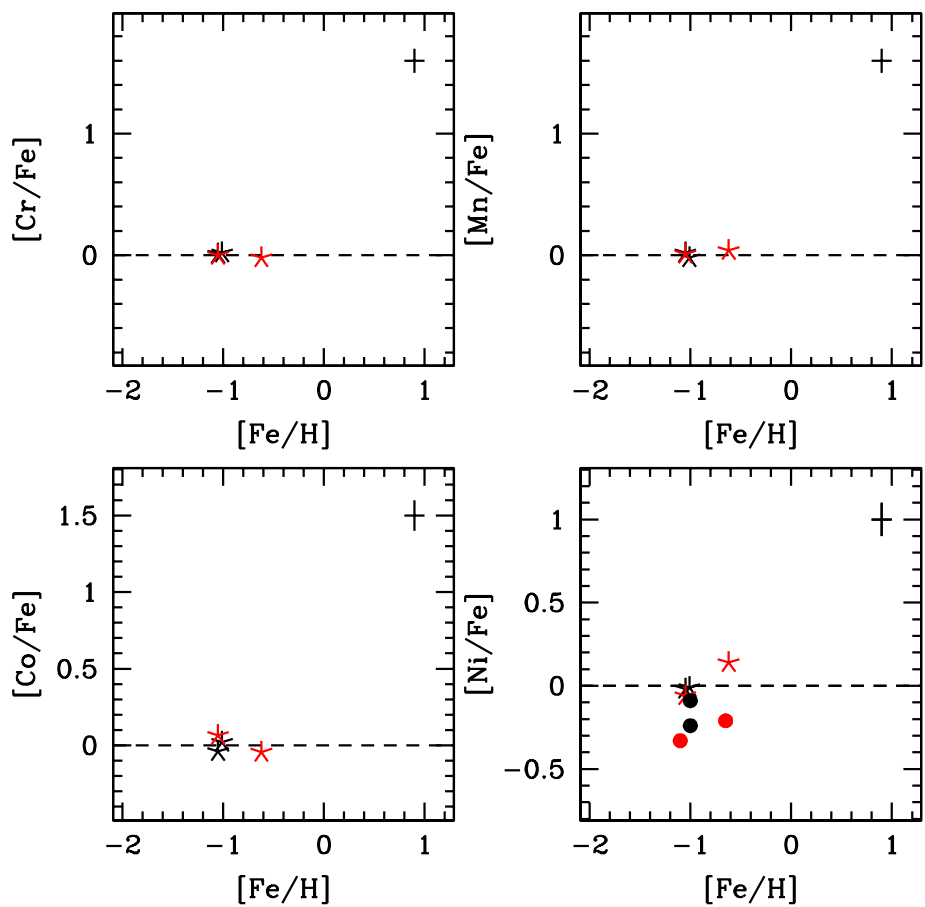}
\caption{The abundance ratios with respect to solar versus
metallicity for the program stars are shown. The red stars
are the mildly H-deficient stars and the black stars are the
normal comparison stars. The filled circles represents
the elemental abundances derived by \citet{johnson10} for
the program stars, the red circles the mildly
H-deficient/He-enhanced stars and black circles the normal
(H-rich) stars. The horizontal dotted lines shows
the solar scaled abundance values. The error bars on the
abundance ratios are shown with crossed thicklines in the top
right corner.
\label{}}
\end{figure}

However, [O/Fe] for our program stars are similar to solar, 
and do not show any depletion/enhancement, and are also 
similar to [O/Fe] $\approx$ $-$0.15 (see figure 11) 
for metal-rich RGBs derived by \citet{johnson10}. 
\citet{marino11} has conducted a high-resolution 
spectroscopic studies for red giant stars of $\omega$ Cen
for deriving Fe, Na, O and n-capture elements. They have 
studied the Na-O anticorrelation for the giants of different 
metallicity. The giants in the metal-rich regime do not show any 
Na-O anti-correlation unlike metal-poor and metal-intermediate giants. 
This is also observed in our program stars.
Lanthanum abundance derived for our program stars is from 
a single line and are in line with the [La/Fe]
vs. [Fe/H] plots given by \citet{johnson10, marino11, marino12, dorazi11}.

For the SGB-a stars, \citet{villanova07} has derived the abundances
for Ca and for Ti along with C, N and Ba. 
The enhancements, [Ca/Fe] and [Ti/Fe] of 0.48 dex and 0.44 dex,
respectively, for SGB-a stars by \citet{villanova07}, are in good 
agreement with the enhancements, [Ca/Fe] and [Ti/Fe] of 0.35 dex 
and 0.3 dex, respectively, for our program stars. Similarly, 
\citet{pancino11} have determined the [$\alpha$/Fe] =$+$0.4 dex, 
[Al/Fe] = $+$0.32 dex, [Fe-peak/Fe] $\sim$ 0.0 dex which are 
in excellent agreement with those determined 
by \citet{villanova07} and also
those determined for metal-rich giants by \citet{johnson10} and
in this study.
These abundance similarities links the SGB-a stars with the
metal rich RGB stars of $\omega$ Cen. 

A very important link between: bMS, SGB-a, metal-rich RGBs,
come from the helium enhancement. An unacceptable lower 
Mg abundance derived for our sample stars, LEID 39048 
and LEID 34225, from the MgH bands, (the 
weaker MgH bands), than that expected for their derived 
stellar parameters, and the Mg abundances from Mg\,{\sc i} lines 
and the uncertainties on these parameters, suggests the lower 
hydrogen/He-enhancement in their atmospheres.

Hence, similar to bMS stars, our sample stars which are metal rich
RGBs, show  mild deficiency in hydrogen or enhanced helium. 
All metal rich RGBs may not be H-poor/He-enhanced, but a 
sub-group of them are. \citet{dupree11} have reported the 
first direct evidence for an enhancement of helium in the 
metal-poor RGBs of $\omega$ Cen by analyzing the near-infrared  
He\,{\sc i} 10830\AA\ transition in about 12 red giants.
From their studies they notice that the He-enhanced
giants show enhanced [Al/Fe] and [Na/Fe], than the He-normal 
giants (see their Figure 10).  Figure 13 shows the 
plot [Al/Fe] vs. [Na/Fe] for our program stars 
along with \citet{dupree11}'s 
sample stars. Our program stars follow the similar trend 
as that of \citet{dupree11}'s sample stars.

\begin{figure}
\epsscale{0.80}
\plotone{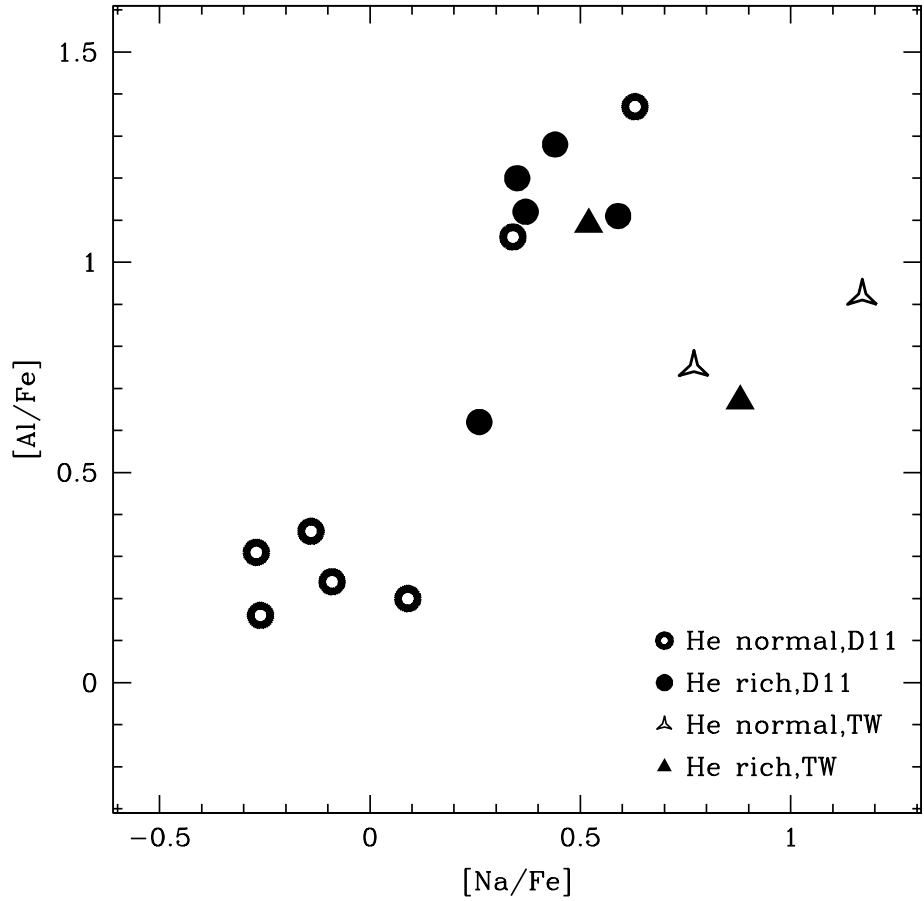}
\caption{The abundance ratios of [Al/Fe] vs. [Na/Fe]
are shown. The open and filled circles are He-normal
and He-enhanced giants, respectively from \citet{dupree11}
and the open and filled triangles are the He-normal and
He-enhanced giants, respectively, from this work (TW).
\label{}}
\end{figure}

 A detailed spectroscopic studies are not available for 
horizontal branch stars. 
However, the helium enhancement in horizontal branch stars
may also be due to helium-flash or any other processes, and may
not be wholly intrinsic \citep{moehler02}.

\section{Conclusions}

{This study based on the evaluation of the strengths of the
MgH bands in the observed high-resolution spectra and 
for the stars' adopted stellar parameters, 
confirms that LEID 39048 and LEID 34225 are 
mildly H-deficient/He-enhanced.
Discovery and the detailed abundance analysis 
of these stars provides a 
direct evidence for the presence of  
He-enhanced metal rich giants in $\omega$ Cen. 
These stars provides crucial link 
to the evolution of the metal rich 
sub-population of MS, sub giant and red giants.}

\acknowledgments
All of the observations reported in this paper were
obtained with the Southern African Large Telescope (SALT).
We thank the anonymous referee for a constructive report
that improved the presentation of this paper.


\begin{thebibliography}{27}
\expandafter\ifx\csname natexlab\endcsname\relax\def\natexlab#1{#1}\fi

\bibitem[{{Alonso} {et~al.}(1999){Alonso}, {Arribas}, \&
  {Mart{\'{\i}}nez-Roger}}]{alonso99}
{Alonso}, A., {Arribas}, S., \& {Mart{\'{\i}}nez-Roger}, C. 1999, \aaps, 140,
  261

\bibitem[{{Anders} \& {Grevesse}(1989)}]{anders89}
{Anders}, E., \& {Grevesse}, N. 1989, \gca, 53, 197

\bibitem[{{Asplund} {et~al.}(2009){Asplund}, {Grevesse}, {Sauval}, \&
  {Scott}}]{asplund09}
{Asplund}, M., {Grevesse}, N., {Sauval}, A.~J., \& {Scott}, P. 2009, \araa, 47,
  481

\bibitem[{{Calamida} {et~al.}(2009){Calamida}, {Bono}, {Stetson}, {Freyhammer},
  {Piersimoni}, {Buonanno}, {Caputo}, {Cassisi}, {Castellani}, {Corsi},
  {Dall'Ora}, {Degl'Innocenti}, {Ferraro}, {Grundahl}, {Hilker}, {Iannicola},
  {Monelli}, {Nonino}, {Patat}, {Pietrinferni}, {Prada Moroni}, {Primas},
  {Pulone}, {Richtler}, {Romaniello}, {Storm}, \& {Walker}}]{calamida09}
{Calamida}, A., {et~al.} 2009, \apj, 706, 1277

\bibitem[{{Da Costa} {et~al.}(2013){Da Costa}, {Norris}, \& {Yong}}]{dacosta13}
{Da Costa}, G.~S., {Norris}, J.~E., \& {Yong}, D. 2013, \apj, 769, 8

\bibitem[{{D'Orazi} {et~al.}(2011){D'Orazi}, {Gratton}, {Pancino}, {Bragaglia},
  {Carretta}, {Lucatello}, \& {Sneden}}]{dorazi11}
{D'Orazi}, V., {Gratton}, R.~G., {Pancino}, E., {Bragaglia}, A., {Carretta},
  E., {Lucatello}, S., \& {Sneden}, C. 2011, \aap, 534, A29

\bibitem[{{Dupree} \& {Avrett}(2013)}]{dupree13}
{Dupree}, A.~K., \& {Avrett}, E.~H. 2013, \apjl, 773, L28

\bibitem[{{Dupree} {et~al.}(2011){Dupree}, {Strader}, \& {Smith}}]{dupree11}
{Dupree}, A.~K., {Strader}, J., \& {Smith}, G.~H. 2011, \apj, 728, 155

\bibitem[{{Hema} \& {Pandey}(2014)}]{hema14}
{Hema}, B.~P., \& {Pandey}, G. 2014, \apjl, 792, L28

\bibitem[{{Hinkle} {et~al.}(2000){Hinkle}, {Wallace}, {Valenti}, \&
  {Harmer}}]{hinkle00}
{Hinkle}, K., {Wallace}, L., {Valenti}, J., \& {Harmer}, D. 2000, {Visible and
  Near Infrared Atlas of the Arcturus Spectrum 3727-9300 A}

\bibitem[{{Johnson} \& {Pilachowski}(2010)}]{johnson10}
{Johnson}, C.~I., \& {Pilachowski}, C.~A. 2010, \apj, 722, 1373

\bibitem[{{Kurucz}(1998)}]{kurucz98}
{Kurucz}, R.~L. 1998, http://kurucz.harvard.edu/

\bibitem[{{Marino} {et~al.}(2011){Marino}, {Milone}, {Piotto}, {Villanova},
  {Gratton}, {D'Antona}, {Anderson}, {Bedin}, {Bellini}, {Cassisi}, {Geisler},
  {Renzini}, \& {Zoccali}}]{marino11}
{Marino}, A.~F., {et~al.} 2011, \apj, 731, 64

\bibitem[{{Marino} {et~al.}(2012){Marino}, {Milone}, {Piotto}, {Cassisi},
  {D'Antona}, {Anderson}, {Aparicio}, {Bedin}, {Renzini}, \&
  {Villanova}}]{marino12}
---. 2012, \apj, 746, 14

\bibitem[{{Mayor} {et~al.}(1997){Mayor}, {Meylan}, {Udry}, {Duquennoy},
  {Andersen}, {Nordstrom}, {Imbert}, {Maurice}, {Prevot}, {Ardeberg}, \&
  {Lindgren}}]{mayor97}
{Mayor}, M., {et~al.} 1997, \aj, 114, 1087

\bibitem[{{McWilliam} \& {Lambert}(1988)}]{mcwilliam88}
{McWilliam}, A., \& {Lambert}, D.~L. 1988, \mnras, 230, 573

\bibitem[{{Moehler} {et~al.}(2002){Moehler}, {Sweigart}, {Landsman}, \&
  {Dreizler}}]{moehler02}
{Moehler}, S., {Sweigart}, A.~V., {Landsman}, W.~B., \& {Dreizler}, S. 2002,
  \aap, 395, 37

\bibitem[{{Pancino} {et~al.}(2011){Pancino}, {Mucciarelli}, {Bonifacio},
  {Monaco}, \& {Sbordone}}]{pancino11}
{Pancino}, E., {Mucciarelli}, A., {Bonifacio}, P., {Monaco}, L., \& {Sbordone},
  L. 2011, \aap, 534, A53

\bibitem[{{Pandey} {et~al.}(2004){Pandey}, {Lambert}, {Rao}, {Gustafsson},
  {Ryde}, \& {Yong}}]{pandey04}
{Pandey}, G., {Lambert}, D.~L., {Rao}, N.~K., {Gustafsson}, B., {Ryde}, N., \&
  {Yong}, D. 2004, \mnras, 353, 143

\bibitem[{{Piotto} {et~al.}(2005){Piotto}, {Villanova}, {Bedin}, {Gratton},
  {Cassisi}, {Momany}, {Recio-Blanco}, {Lucatello}, {Anderson}, {King},
  {Pietrinferni}, \& {Carraro}}]{piotto05}
{Piotto}, G., {et~al.} 2005, \apj, 621, 777

\bibitem[{{Ram{\'{\i}}rez} \& {Allende Prieto}(2011)}]{ramirez11}
{Ram{\'{\i}}rez}, I., \& {Allende Prieto}, C. 2011, \apj, 743, 135

\bibitem[{{Simpson} \& {Cottrell}(2013)}]{simpson13}
{Simpson}, J.~D., \& {Cottrell}, P.~L. 2013, \mnras, 433, 1892

\bibitem[{{Sneden}(1973)}]{sneden73}
{Sneden}, C.~A. 1973, PhD thesis, THE UNIVERSITY OF TEXAS AT AUSTIN.

\bibitem[{{Sollima} {et~al.}(2005){Sollima}, {Ferraro}, {Pancino}, \&
  {Bellazzini}}]{sollima05}
{Sollima}, A., {Ferraro}, F.~R., {Pancino}, E., \& {Bellazzini}, M. 2005,
  \mnras, 357, 265

\bibitem[{{Sumangala Rao} {et~al.}(2011){Sumangala Rao}, {Pandey}, {Lambert},
  \& {Giridhar}}]{sumangala11}
{Sumangala Rao}, S., {Pandey}, G., {Lambert}, D.~L., \& {Giridhar}, S. 2011,
  \apjl, 737, L7

\bibitem[{{Tailo} {et~al.}(2016){Tailo}, {Di Criscienzo}, {D'Antona}, {Caloi},
  \& {Ventura}}]{tailo16}
{Tailo}, M., {Di Criscienzo}, M., {D'Antona}, F., {Caloi}, V., \& {Ventura}, P.
  2016, \mnras, 457, 4525

\bibitem[{{Villanova} {et~al.}(2007){Villanova}, {Piotto}, {King}, {Anderson},
  {Bedin}, {Gratton}, {Cassisi}, {Momany}, {Bellini}, {Cool}, {Recio-Blanco},
  \& {Renzini}}]{villanova07}
{Villanova}, S., {et~al.} 2007, \apj, 663, 296

\end{thebibliography}


\end{document}